\documentclass[10pt,journal,compsoc]{IEEEtran}
\usepackage[nocompress]{cite}

\usepackage{graphicx}
\graphicspath{{./Images/}}
\DeclareGraphicsExtensions{.pdf,.jpg}

\usepackage[cmex10]{amsmath}
\usepackage[caption=false,font=footnotesize,labelfont=sf,textfont=sf]{subfig}
\usepackage{fixltx2e}
\usepackage{url}

\usepackage{todonotes}

\begin{document}

\title{A Visible Light-based Positioning System}

%
%

\author{Yiqing~Hu,~\IEEEmembership{Student Member,~IEEE,}
        Yan~Xiong,
        Wenchao~Huang,
        Xiang-Yang~Li,~\IEEEmembership{Fellow,~IEEE,}
        Yanan~Zhang,
        Xufei~Mao,
        Panlong~Yang,
        and~Caimei~Wang
\IEEEcompsocitemizethanks{\IEEEcompsocthanksitem Y. Hu, Y. Xiong, W. Huang and C. Wang is with the Department of Computer Science, University of Science and Technology of China, Hefei, China, 230026.\protect\\
E-mail: huyiqing@mail.ustc.edu.cn, \{yxiong,huangwc8\}@ustc.edu.cn, wangcmo@mail.ustc.edu.cn\protect\\
X.Li is with Department of Computer Science, Illinois Institute of Technology, Chicago, USA.\protect\\
E-mail: xli@cs.iit.edu\protect\\
Y.Zhang and X.Mao is with Department of Software Engineering, and TNLIST, Tsinghua University, Beijing, China.\protect\\
E-mail: \{zhangyanan1230, xufei.mao\}@gmail.com\protect\\
P.Yang is with Institute of Communication Engineering, PLAUST, Nanjing, China.\protect\\
E-mail: panlongyang@gmail.com}}

\IEEEtitleabstractindextext{%
\begin{abstract}
In this paper, we propose a novel indoor localization scheme, Lightitude, by exploiting ubiquitous visible lights, which are necessarily and densely deployed in almost all indoor environments.
We unveil two phenomena of lights available for positioning: 1) the light strength varies according to different light sources, which can be easily detected by light sensors embedded in COTS devices (e.g., smart-phone, smart-glass and smart-watch);
2) the light strength is stable in different times of the day thus exploiting it can avoid frequent site-survey and database maintenance. Hence, a user could locate oneself by differentiating the light source of received light strength (RLS).
However, different from existing positioning systems that exploit special LEDs, ubiquitous visible lights lack fingerprints that can uniquely identify the light source, which results in an ambiguity problem that an RLS may correspond to multiple positions.
Moreover, RLS is not only determined by device's position, but also seriously affected by its orientation, which causes great complexity in site-survey.
To address these challenges, we first propose and validate a realistic light strength model that can attributes RLS to arbitrary positions with heterogenous orientations.
This model is further perfected by taking account of the device diversity, influence of multiple light sources and shading of obstacles.
Then we design a localizing scheme that harness user's mobility to generate spatial-related RLS to tackle the position-ambiguity problem of a single RLS, which is robust against sunlight interference, shading effect of human-body and unpredictable behaviours (e.g., put the device in pocket) of user.
Experiment results show that Lightitude achieves mean accuracy $1.93$m and $1.98$m in office ($720m^2$) and library scenario ($960m^2$) respectively.
\end{abstract}
}

\IEEEdisplaynontitleabstractindextext
\IEEEpeerreviewmaketitle


\maketitle

\IEEEraisesectionheading{\section{Introduction}\label{sec:introduction}}
Indoor localization is crucial for location based services.
Ever-increasing demands by retailers, airports and shopping centers give credit to the indoor localization industry, which is promising to grow to $\$5$ billion by 2018~\cite{connolly2013}.
Despite the strong demand, there exist few approaches that perform satisfactorily in every scenario due to their deployment restrictions.
We envision a scenario as an example: in a big mall, retailers intend to deliver advertisements of new goods to nearby potential customers through the pre-installed APP in user's smart device.
In this scenario, mainstream WiFi-based positioning approaches~\cite{youssef2005horus,joshi2013pinpoint} are stumbling due to the limited number of WiFi access points (APs).
We conducted a comprehensive site-survey in 14 big indoor public areas (railway stations, malls, supermarkets, hospitals, etc.) in downtown, and find that the AP density is about one AP per $363 m^2$, which may be sufficient for communication, but is far from enough for accurate localization.

\begin{figure}[t]
  \centering
  \subfloat[RLS distribution]{
    \includegraphics[width=0.24\textwidth]{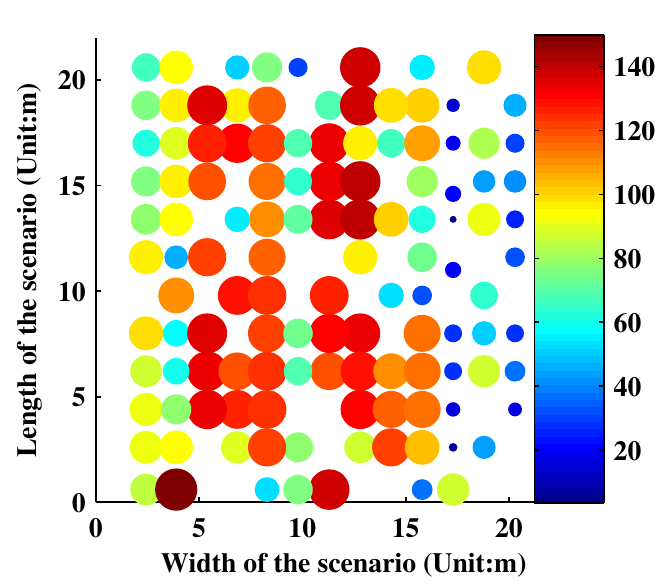}
    \label{fig:2:a}}
    \hfil
  \subfloat[Solution space]{
    \includegraphics[width=0.22\textwidth]{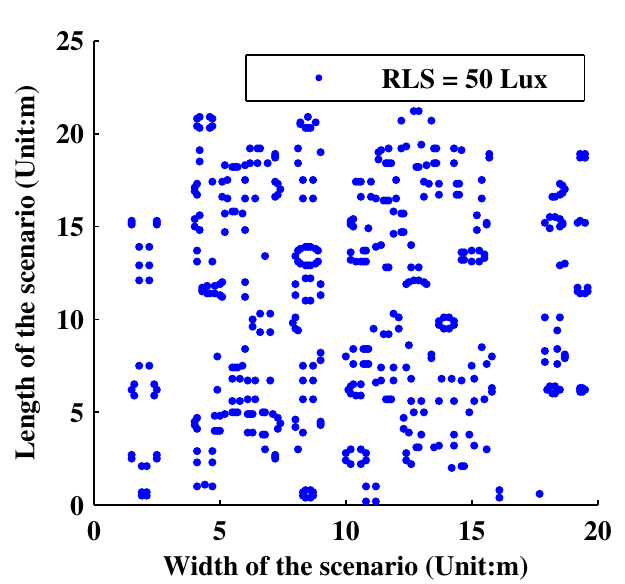}
    \label{fig:2:b}}
  \caption{(a) shows a obvious and distinguishable RLS distribution at altitude $1$m in one experiment scenario.
  Each spot indicates RLS sampled under a light source.
  (b) shows the solution space for a specified RLS. An RLS is correspond to multiple positions.}
  \label{fig:2}
\end{figure}

Compared with the limited number of APs, existing illumination infrastructures in these public areas facilitate approaches that exploit visible lights.
In this paper, we propose a novel indoor localization scheme, named Lightitude, which exploits ubiquitous visible lights that are necessarily and densely deployed in almost all indoor environments.
It is based on our new finding: the light strength varies according to light sources in different locations, and the difference is unexpectedly obvious for light sensors, as shown in Fig~\ref{fig:2:a}.
Different from current visible light-based positioning (VLP) systems~\cite{rajagopal2014visual,liepsilon,luxapose2014}, Lightitude does not require redeploying infrastructures or special devices (e.g., customized LEDs / light sensor with high sample rate) before localization.
Due to the ubiquity and zero-cost of Lightitude, it can be directly applied as an auxiliary subsystem coexisting with WiFi-based schemes or even an independent positioning system.
Furthermore, compared with WiFi signals, the light strength is stable in different times of the day.
Hence, Lightitude avoids frequent site-survey and database maintenance.

Though the characteristic that light strength can infer the locations is similar to that of WiFi fingerprint-based schemes,
we face several new challenges in exploiting the ubiquitous visible lights.
\textbf{Firstly}, RLS is easily influenced by receiving device's orientation and altitude.
Even a subtle status change of the receiving device will cause huge RLS deviation, even keeping the device at a fixed coordinate, which further causes fingerprint mismatch.
Hence, it's hard to leverage traditional fingerprints site-survey (2D position-traversing) to link RLS with position information.
At the same time, Traversing a six-dimensional variable space (roll, pitch, yaw and 3D coordinate of the receiving device) to collect fingerprints may solve this problem, but this unfortunately would be extremely labor intensive and not practical.
\textbf{Secondly}, a unique RLS value may correspond to multiple possible positions.
The key reason is that, compared with WiFi, RLS is merely a scalar that lacks unique identification such as light source's ID.
As a result, similar RLS can be captured at multiple locations, which implies different possible locations for the device, as shown in Fig~\ref{fig:2:b}.
The drawbacks of fingerprint-based solution motivate us to design a light strength model to replace fingerprint-collection.

Facing the first dilemma, we propose and rely on a light strength model to connect every receiving device's status (one item in the six-dimensional variable space) with RLS. It replaces the theoretical light strength model~\cite{lcl}, which is inapplicable in our scenario due to non-ideal light source and receiving device.
Specifically, we investigate the impact of radiation angle, incidence angle and relative distance from the device to the corresponding light source on RLS, and then design a light strength model.
Using this model, we could calculate RLS at an arbitrary status of the receiving device in the six-dimensional variable space.
Additionally, we also take account of \textbf{device diversity}, \textbf{influence of multiple light sources} and \textbf{shading of obstacles}, and finally form a light strength model that works well in real scenario.

To address the second challenge, we harness user's mobility to capture a spatially related RLS set, then incorporate it and its corresponding inertial sensor data into a particle filter for positioning.
We estimate the user's location without assuming a priori knowledge about user-specific characteristics like stride length and heading.
The insight is that RLS set collected by a user serves as a filter.
A deviated stride length and heading generate a candidate trace whose corresponding RLS set is unmatched with what the user has collected.
As a result, the biased candidate trace is eliminated.
To make the localizing scheme robust, we further design a mechanism that takes advantage of \textbf{sunlight interference} rather than simply eliminates its impact.
We also show Lightitude's robustness under the \textbf{shading effect of human-body}.
Furthermore, encountering \textbf{unpredictable behaviours} of users like picking a phone call or putting the device in pocket, we design mechanisms to detect and eliminate their impact by comprehensive analysis.
At last, we \textbf{coexist the localizing scheme with WiFi} to accelerate its convergence speed and improve its positioning accuracy.

We built a prototype of Lightitude, and evaluated it in an office about $720 m^2$ with $39$ common fluorescent lamps, and one floor in the school library about $960 m^2$ with $123$ common fluorescent lamps.
We used a Google Nexus $4$, a Google Nexus $7$ and a smart watch Moto $360$ as the receiving devices in the experiment.
Our experiment has the following results: compared with RLS sampled by the light sensor, the error of the light strength model is limited within $10$ Lux in $50$ percentages, and $50$ Lux in $95$ percentages.
Examining by real traces with a total length of 6.04km in the office and 14.9km in the library, Lightitude yields mean accuracy $1.93$m and $1.98$m in these two scenarios respectively.
Specially in the big library scenario, through walking a longer distance for $15$ steps and $20$ steps by the user, Lightitude itself achieves mean positioning accuracy $1.79$m, $1.26$m respectively, which is sufficient enough to precisely locate user even between target shelves.
Lightitude performs well even with the shading effect of obstacles, unpredictable behaviours of users, interference of sunlight and human-body's shading effect.
Our approach provides a new perspective on how to leverage the most ubiquitous visible lights.

The rest of the paper is organized as follows.
In Section~\ref{sec:relatedwork}, we discuss the state-of-the-art indoor localization techniques.
In Section~\ref{sec:overview}, we present the system overview of Lightitude.
In Section~\ref{sec:lightModel}, we introduce our light strength model, then propose the localizing scheme in Section~\ref{sec:localizingmodule}, followed with the experiments in Section~\ref{sec:experiment}.
Section~\ref{sec:conclusion} concludes the paper.

\section{Related Work}
\label{sec:relatedwork}
\subsection{Recent Localization Schemes}
WiFi-based positioning system is the current mainstream localization solution~\cite{bahl2000radar,youssef2005horus,adib2013see,joshi2013pinpoint,adib20133d,rai2012zee,yang2012locating,shen2013walkie,wang2012no}. RSS fingerprint-based schemes~\cite{bahl2000radar,youssef2005horus} rely on a centralized localization service like a database, and could achieve meter level accuracy.
Approaches that based on multiple antennas~\cite{adib2013see,joshi2013pinpoint,adib20133d} achieve sub-meter level accuracy, but it's difficult to deploy these customized APs in scale.
By engaging user's motion, LiFS~\cite{yang2012locating} avoids the labor cost in building a centralized localization service, at the cost of accuracy loss at some extent.
Landmark approaches~\cite{shen2013walkie,wang2012no} work well in narrow spaces like corridor, but they are suffering in the large space environment due to the difficulty in abstracting landmarks.
Methods based on other mediums like acoustic~\cite{liu2012push,peng2007beepbeep,DBLP:conf/infocom/HuangXLLMYL14}, FM~\cite{chen2012fm} and RFID~\cite{wang2013dude} can achieve meter and even sub-meter level accuracy, but these schemes have their own deployment limitations and constraints of application scenario.

\subsection{Visible Light-based Positioning}
Many VLP schemes are proposed in recent years.
In~\cite{liu2008basic,randall2007luxtrace}, customized fluorescent lamps work as landmarks.
Leveraging customized fluorescent lamps also, in FiatLux~\cite{5433479}, light fingerprint database is built to achieve room-level accuracy.
Epsilon~\cite{liepsilon} exploits a customized light sensor to capture signals sent by pre-modulated LEDs, and could achieve sub-meter level accuracy.
In this scheme, both the light sending and receiving terminal are customized. In~\cite{yoshino2008high,rajagopal2014visual,luxapose2014}, image sensor is exploited to capture messages and beacons sent from LED lamps, leveraging the rolling shutter effect.
All these schemes require customized devices.

In Lightitude, both the light sources (existing fluorescent lamps) and receiving devices are COTS devices, thus no additional infrastructures are required. Moreover, Lightitude can work on pervasive light sources like customized LEDs as well, by attributing different light strength characteristics to different LEDs using pre-modulate technique.

\section{System overview}
\label{sec:overview}
\begin{figure}[h]
  \centering
  \includegraphics[width=0.45\textwidth]{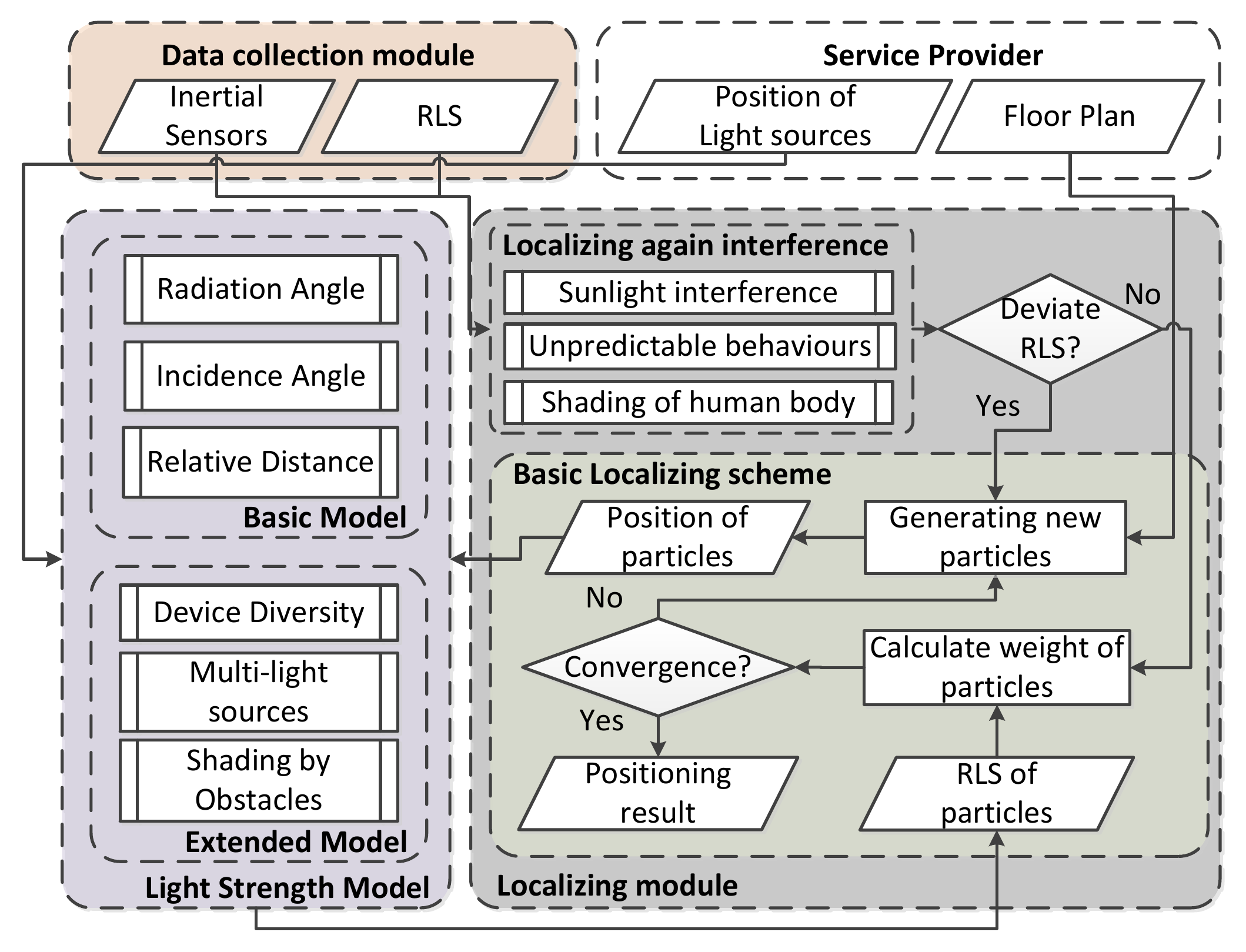}
  \caption{Lightitude System Architecture. }
  \label{fig:1}
\end{figure}

Fig~\ref{fig:1} shows the system architecture of Lightitude.
Lightitude consists of three parts: data collection, light strength model, and localizing scheme.
The last two parts are key components of Lightitude.
We describe the flow of operation here, and expand on the technical details in the following sections.

\textbf{Light Strength Model: }The key functionality of the light strength model is to calculate the weights of candidate particles (positions), which indicate their potential to be the ground truth position of the user.
A particle obtains a big weight when its RLS is close to current collected RLS, and vice versa.

The major challenge is that the previous classical model, i.e., model proposed in Epsilon, cannot be directly applied in our localizing scheme, because the devices we exploit are different from the customized ones specified in Epsilon.
To tackle this problem, we investigate the influence factors of RLS: light radiation angle $\phi$, light incidence angle $\theta$ and distance from the lamp to the receiving device $d$.
Basing on the investigation, we build a light strength model of a single light source first, then expand it to make it capable in complex scenarios.

\textbf{Localizing Scheme: }In the localization module, we design a particle filter to translate the pedometer's result (step number) into distance, meanwhile avoid assuming a priori knowledge about user-specific knowledge like stride length and heading.
In the pedometer module, we employ a local variance threshold method~\cite{jimenez2009comparison} to count steps.
The key idea underlying this solution is that, starting from a particle with known position, the relationship between consecutive ones provides not only candidate positions, but also possible stride length and heading.
Particles with biased stride length and heading are eliminated on account of their deviated RLS.
User's position, stride length or heading converge simultaneously after enough iterations.

These components work together to ensure the user can locate oneself by taking a few steps.
In the next section, we first introduce the characteristic of RLS.

\section{Light Strength Model}
\label{sec:lightModel}
In indoor environments, the light strength distribution is dominated by strong light sources (e.g., fluorescent lamps, LEDs), and it's easy to find dozens of them in big scenarios (e.g., there are $123$ lamps in our library scenario).
Light sensor integrated in most smart devices can capture RLS, whose value is determined by nearby strong light sources.

\begin{figure}[b]
  \centering
  \subfloat[Different lamps' characteristics]{
    \includegraphics[width=0.23\textwidth]{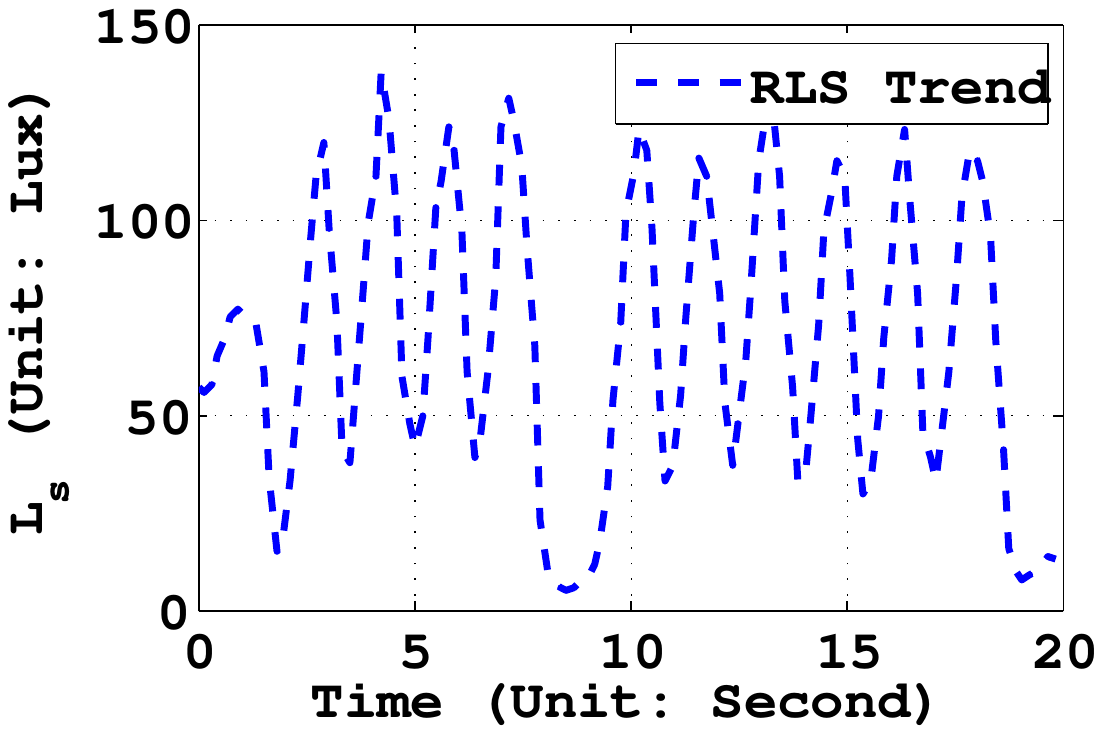}
    \label{fig:3:a}}
    \hfil
  \subfloat[WiFi RSS in half an hour]{
    \includegraphics[width=0.23\textwidth]{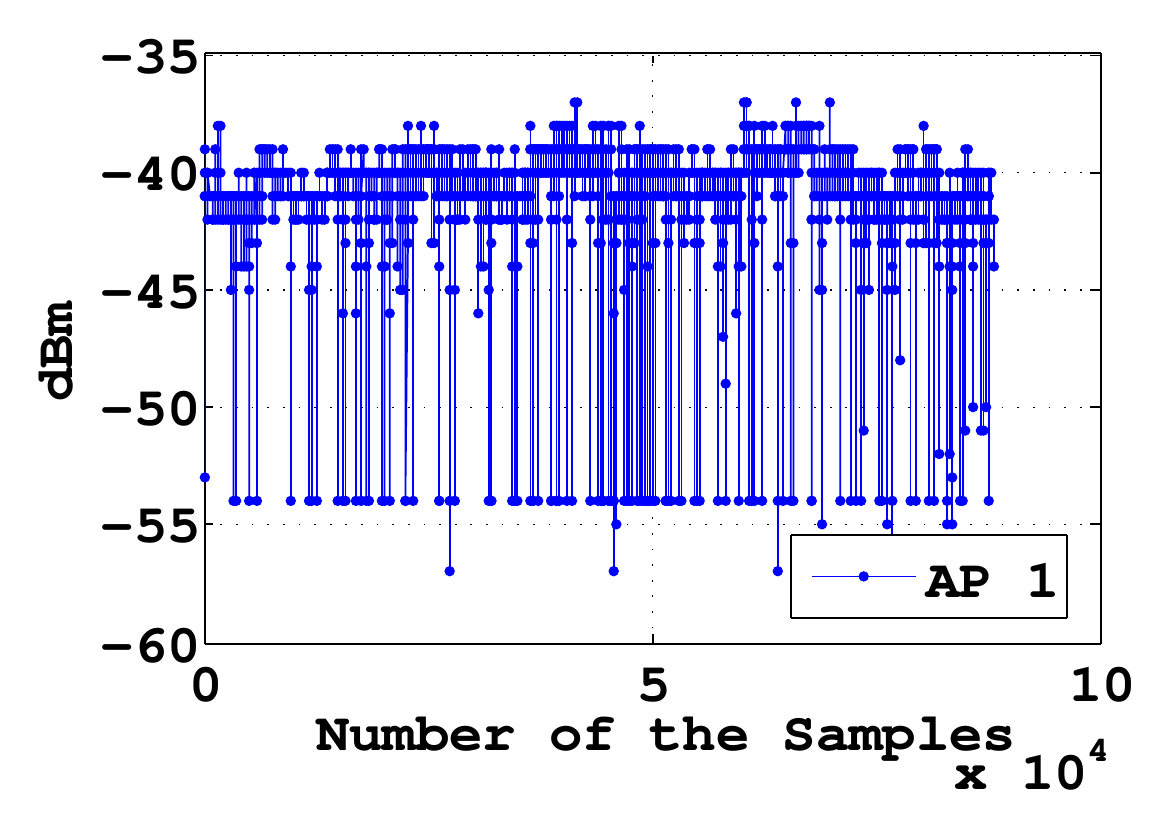}
    \label{fig:3:b}}
    \hfil
  \subfloat[RLS in different times of the day]{
    \includegraphics[width=0.23\textwidth]{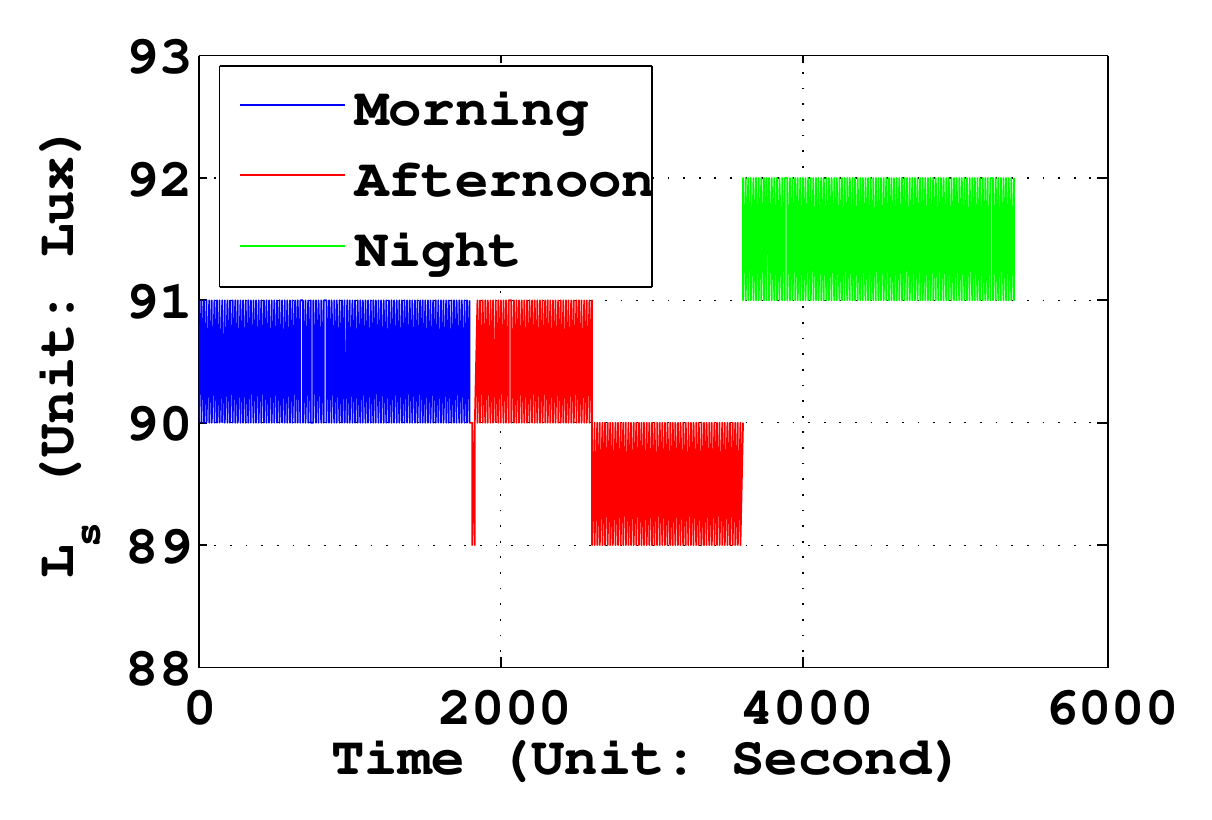}
    \label{fig:3:c}}
    \hfil
  \subfloat[Roll, Pitch, Yaw]{
    \includegraphics[width=0.23\textwidth]{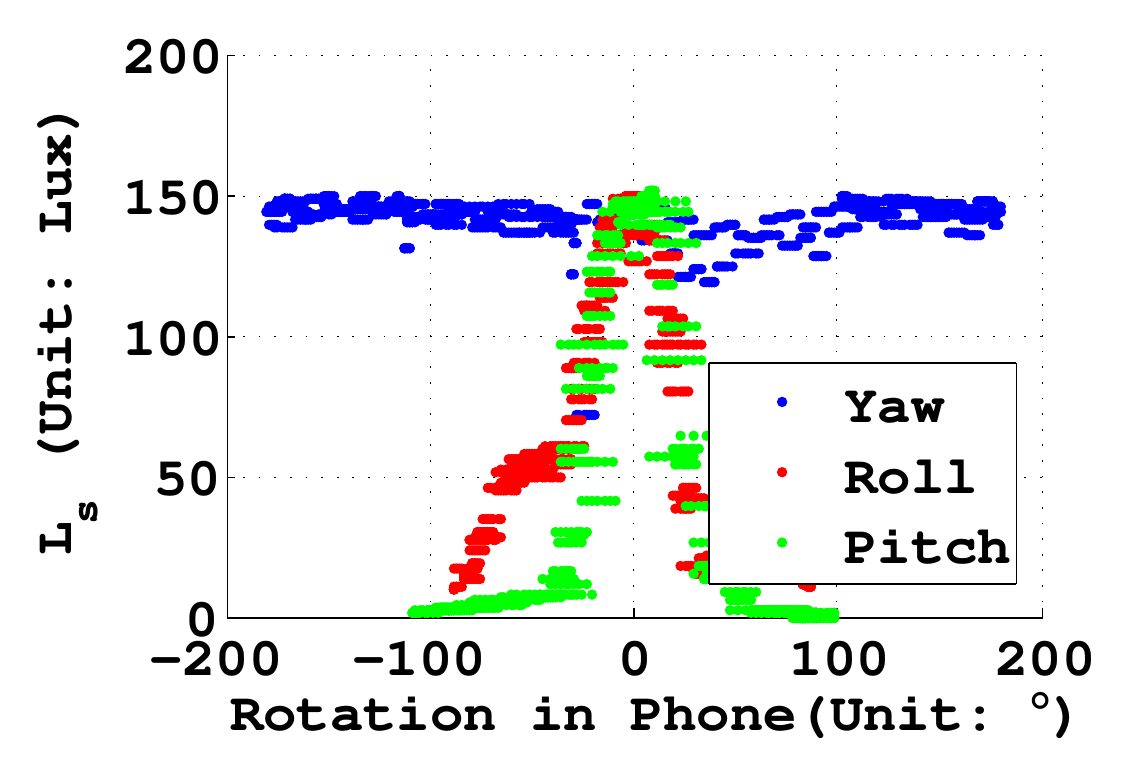}
    \label{fig:3:d}}
  \caption{(a): a RLS trend formed by a volunteer walks across $11$ lamps at different times of the day.
  (b): WiFi RSS fluctuation in half an hour. The receiving device is kept fixed on a desk.
  (c): RLS fluctuation at different times of the day. The position of the receiving device is same as that in (b). Data were collected at $9$ a.m., $3$ p.m. and $9$ p.m. respectively, each for half an hour.
  (d): the rotation of the receiving device influences RLS. }
  \label{fig:3}
\end{figure}

\textit{We find that RLS is associated with position information.}
Keeping the receiving device facing upright and altitude fixed, a volunteer walks across several fluorescent lamps.
We depict this RLS trend collected by a volunteer in Fig~\ref{fig:3:a}.
A RLS trend with several peaks can be captured in the volunteer's route, and each RLS peak is associated with a lamp's $2$D position.
Moreover, strength of these peaks are different and distinguishable, as depicted in Fig~\ref{fig:3:a}.

\textit{We also find an advantage that the RLS is more stable compared with WiFi RSS.}
We can see in Fig~\ref{fig:3:b} and Fig~\ref{fig:3:c} that the WiFi RSS fluctuates obviously, while RLS varies only slightly.

Despite that the relation between RLS and its corresponding position information is obvious and stable in different times of the day, we still find it formidable to exploit this relation for positioning directly.
The crucial reason is that, compared with other received signal strengths (e.g., WiFi signal, magnetic field strength), RLS is greatly influenced by rotation of the receiving device.
As shown in Fig~\ref{fig:3:d}, roll and pitch of the device have a great impact on RLS, even keeping the device at a fixed coordinate.
As a result, the intuitive fingerprinting method works only by forcing the user holds the device at a fixed altitude and keeps it facing upright all the time.
By this means, we can reduce the six-dimensional variable space (roll, pitch, yaw and $3$D coordinate of the receiving device) to a two-dimensional variable space ($2$D coordinate of the receiving device).
Unfortunately, it is difficult to put such strong restrictions on user's motion.
Facing this dilemma, we propose a light strength model that depends on the six-dimensional variable space.
This model links every status of the receiving device with a corresponding RLS.

\begin{figure}[b]
  \centering
  \subfloat[Influencing Factors]{
    \includegraphics[width=0.15\textwidth]{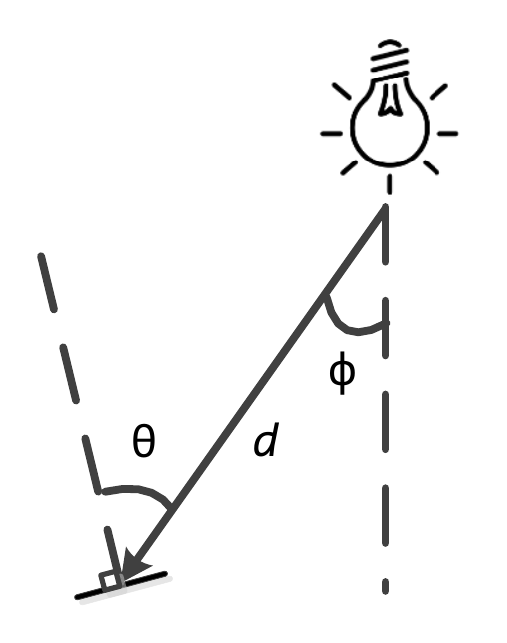}
    \label{fig:4:a}}
    \hfil
  \subfloat[Coordinate System Transforming]{
    \includegraphics[width=0.28\textwidth]{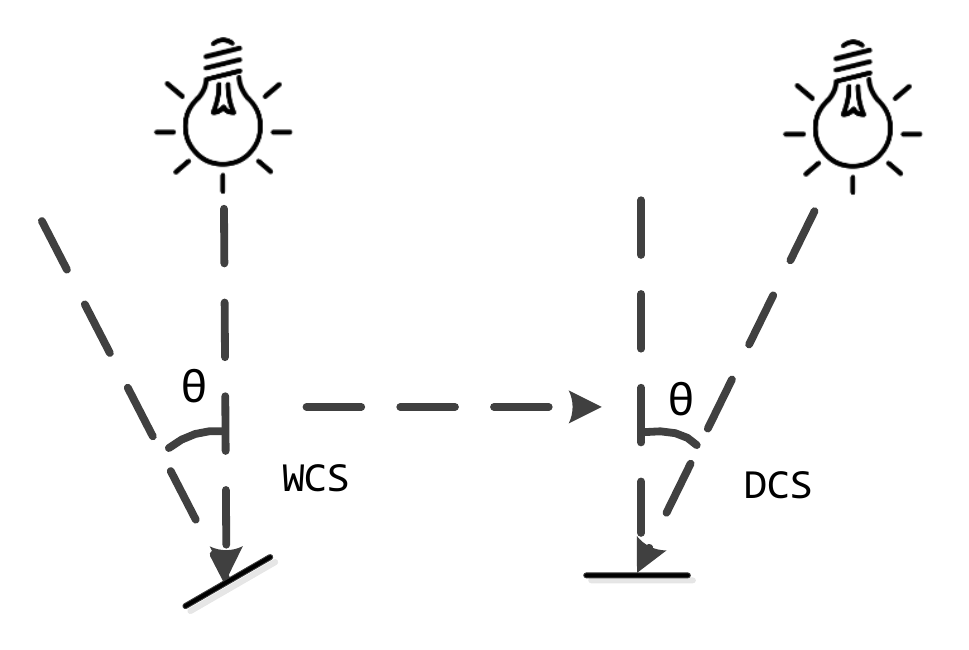}
    \label{fig:4:b}}
  \caption{(a) demonstrates three influencing factors of RLS respectively: radiation angle $\phi$, incidence angle $\theta$ and distance from light source to receiving device $d$.
  (b) demonstrates the transformation from the world coordinate system (WCS) to the device coordinate system (DCS).}
  \label{fig:4}
\end{figure}

To build the light strength model, we start to analyze the influence factors of RLS.
The basic influence factors of RLS contains: light radiation angle $\phi$, light incidence angle $\theta$ and distance from the lamp to the receiving device $d$, as shown in Fig~\ref{fig:4:a}.
However, several other influence factors like device diversity, multi-light source scenario and shading by fixed obstacles should be taken into consideration for applicability of the model.
In this section, we first introduce the basic model, then extend it to make it applicability in real scenario.

\begin{figure*}[t]
  \centering
  \subfloat[$\phi$ and $L_s$]{
    \includegraphics[width=0.22\textwidth]{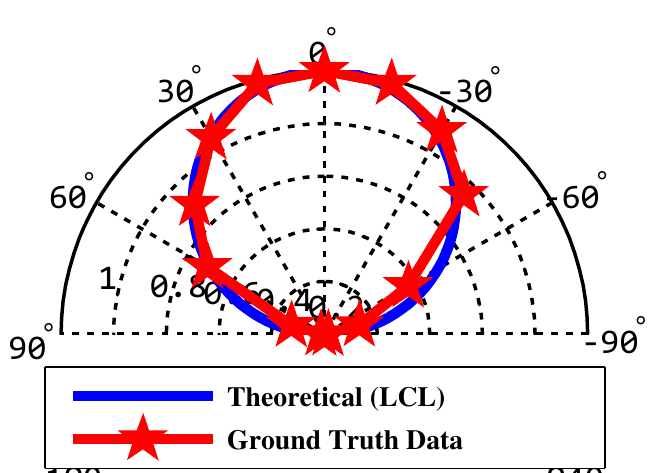}
    \label{fig:5:a}}
    \hfil
  \subfloat[$\theta$ and $L_s$]{
    \includegraphics[width=0.22\textwidth]{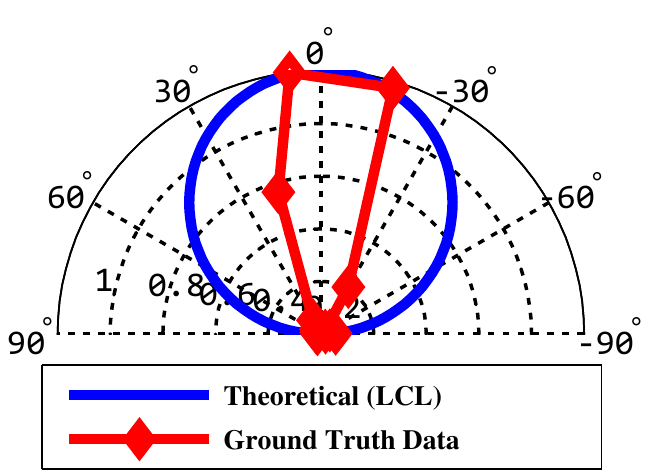}
    \label{fig:5:b}}
    \hfil
  \subfloat[$d$ and $L_s$]{
    \includegraphics[width=0.22\textwidth]{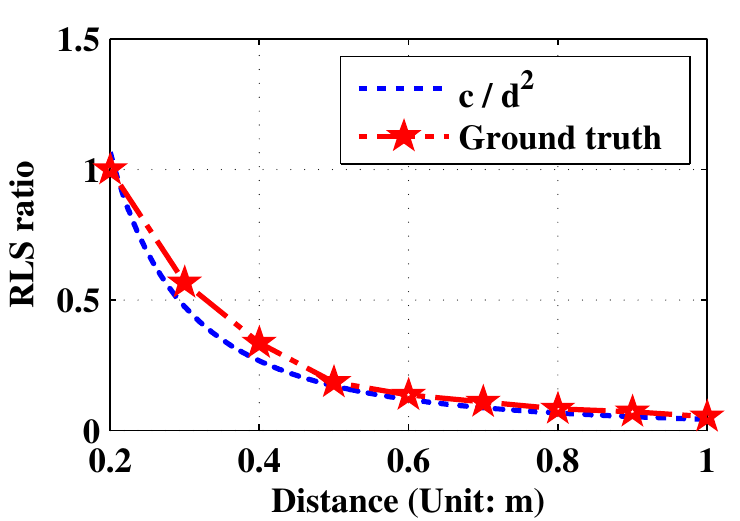}
    \label{fig:5:c}}
    \hfil
  \subfloat[Gaussian fit: $\theta$ and $L_s$]{
    \includegraphics[width=0.22\textwidth]{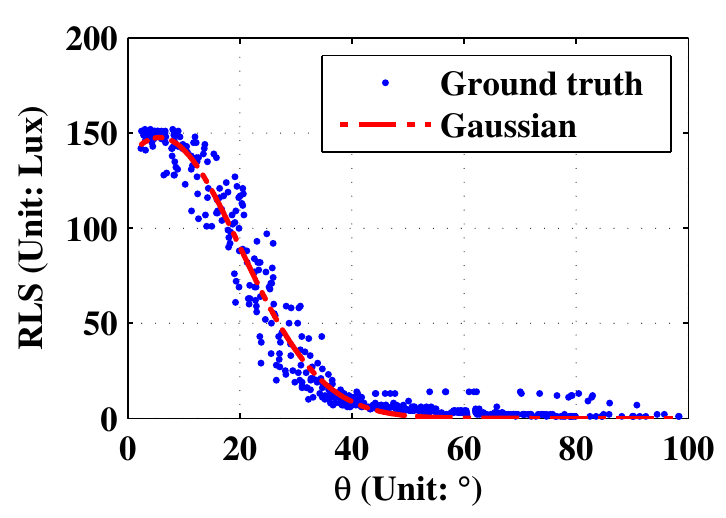}
    \label{fig:5:d}}
  \caption{The relationship between ($\phi$, $L_s$), ($\theta$, $L_s$) and ($d$, $L_s$). (d) shows Gaussian function performs well in describing the function between $\theta$ and $L_s$.}
  \label{fig:5}
\end{figure*}

\subsection{Basic Model}
\label{sec:basicmodel}
Despite that previous work Epsilon~\cite{liepsilon} introduces the basic influence factors $\phi$, $\theta$ and $d$ and formulates them, we still find it inappropriate to directly apply their results in our scheme.
The crucial reason is that the devices we exploit are different from the customized ones specified in Epsilon.
Light sensor integrated in most commodity smart devices is not directly exposed to visible lights, due to shading of device's hull.
Therefore, we reinvestigate these influence factors to design a light strength model.

\textbf{Radiation Angle $\phi$: }We use theoretical model Lambert Cosine Law (LCL)~\cite{lcl} to model the relation between radiation angle $\phi$ and RLS.
By assuming a fixed zero point in the scenario, the receiving device has a $3$D coordinate in the world coordinate system (WCS).
Assume that a volunteer holds the receiving device at $( X_{i}, Y_{i}, Z_{i} )$, and the closest lamp's coordinate is $( l_{x_i}, l_{y_i}, l_{z_i} )$, the orientation vector from the lamp to the receiving device is $v_w = ( l_{x_i} - X_{i}, l_{y_i} - Y_{i}, l_{z_i} - Z_{i} )$.
Then we can get radiation angle $\phi$ by equation~\ref{eqn:phi}.

\begin{equation}
\phi = \arccos(\frac{l_{z_i} - Z_{i}}{\sqrt{(l_{x_i} - X_{i}) ^ 2 + (l_{y_i} - Y_{i}) ^ 2 + (l_{z_i} - Z_{i}) ^ 2}})
\label{eqn:phi}
\end{equation}

In LCL, the luminous intensity observed from a diffuse radiator is directly proportional to the cosine of the radiation angle $\phi$.
Here we adopt LCL to model the relationship between $\phi$ and RLS.
Setting the distance from the receiving device to lamp $d$ and incidence angle $\theta$ fixed, we sampled RLS with $\phi$ ranges from $-90^\circ$ to $90^\circ$, with step size $15^\circ$.
As shown in Fig~\ref{fig:5:a}, LCL performs well in fitting the relation between RLS and $\phi$.
In equation~\ref{eqn:lcl}, $L_s$ indicates RLS and $L_0$ is the maximum RLS received at a fixed distance from the light source to the device.
$L_0$ can be achieved when receiving device's facing orientation vector is antiparallel with $v_w$.

\begin{equation}
L_s = L_0 \cdot \cos(\phi)
\label{eqn:lcl}
\end{equation}

\textbf{Incidence Angle $\theta$:} Similar to Epsilon~\cite{liepsilon}, we try to use cosine rule in equation~\ref{eqn:lcl} to model the relation between incidence angle $\theta$ and RLS.
Changes in roll, pitch and yaw of the receiving device will all influence incidence angle $\theta$, so we use $\theta$ in device coordinate system (DCS) to cover these three factors.
$v_d = ( p_x, p_y, p_z )$ is the orientation vector from the light source to the receiving device in DCS.
It derives from $v_w$ by pre-multiplying receiving device's rotation vector.
The coordinate system transforming is shown in Fig~\ref{fig:5:b}.
We use equation~\ref{eqn:angleinphone} to calculate $\theta$.

\begin{equation}
\cos(\theta) = \frac{p_z}{\sqrt{p_x^2+p_y^2+p_z^2}}
\label{eqn:angleinphone}
\end{equation}

However, experiment shows that cosine rule in equation~\ref{eqn:lcl} performs unsatisfactorily on $\theta$, as shown in Fig~\ref{fig:5:b}.
The phenomenon still exists when using other receiving devices.
Cosine rule works well on customized receiving device in Epsilon~\cite{liepsilon}, because another special light sensor is integrated into the receiving device through the audio jack.
So this light sensor is directly exposed to visible light, and will not be shielded by hull of the receiving device.
Therefore, using COTS devices as receiving terminals, cosine rule is inaccurate in measuring the relationship between $\theta$ and RLS.

To accurately model the rule between $\theta$ and RLS, we conduct experiment to find their inherent relationship.
Setting the distance from the receiving device to lamp $d$ and radiation angle $\phi$ fixed, we change $\theta$ from $-90^\circ$ to $90^\circ$ to collect ground truth data.
We find the Gaussian function performs well in fitting the ground truth data, as shown in Fig~\ref{fig:5:d}.
So we use Gaussian function to describe the relation between $\theta$ and RLS, as shown in equation~\ref{eqn:theta}.
$\delta$ and $\theta_0$ are constants, and we empirically set $\delta = 0.3$ and $\theta_0 = 0^\circ$.

\begin{equation}
L_s = L_0 \cdot e^{-(\frac{\theta - \theta_0}{\delta})^2}
\label{eqn:theta}
\end{equation}

\textbf{Distance from Light Source to Device $d$: } Since distance from the light source to the device also influences RLS, we model their relationship to complement our light strength model.
Inverse-square law is widely used in describing the energy-decay phenomenon.
Hence, we apply it to describe the relation between RLS and $d$.
Fixing both radiation angle $\phi$ and incidence angle $\theta$ at $0^\circ$, we vary $d$ from $0.2$m to $1$m with step size $0.1$m.
We use $\frac{C}{d^2}$ to fit collected ground truth data, in which $C$ is a constant.
As shown in Fig~\ref{fig:5:c}, inverse-square law accurately describes the relation between RLS and $d$, with RMSE $0.04911$.

A basic light strength model can be built by combining all three factors $\phi$, $\theta$, $d$ together:

\begin{equation}
L_s = L_0 \cdot \frac{\cos(\phi) \cdot e^{-(\frac{\theta - \theta_0}{\delta})^2}}{d ^ 2}
\label{eqn:rlsall}
\end{equation}

\subsection{Extended Model}
\label{sec:extendedmodel}
Beside the basic model that considers only the geometric relationship between light source and receiving device, in this subsection, we discuss several other influence factors of RLS, which are even more severe factors in real applications.
Considering the device diversity, influence of multiple light sources and the shield by obstacles, we extend the basic model and make it applicable in real scenarios.

\begin{figure*}[t]
  \centering
  \subfloat[RLS of devices]{
    \includegraphics[width=0.21\textwidth]{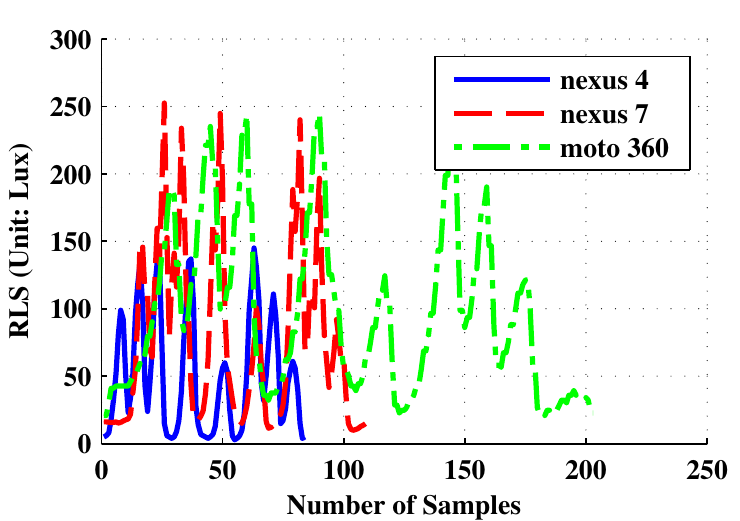}
    \label{fig:6:a}}
    \hfil
  \subfloat[RLS Ratio of devices]{
    \includegraphics[width=0.21\textwidth]{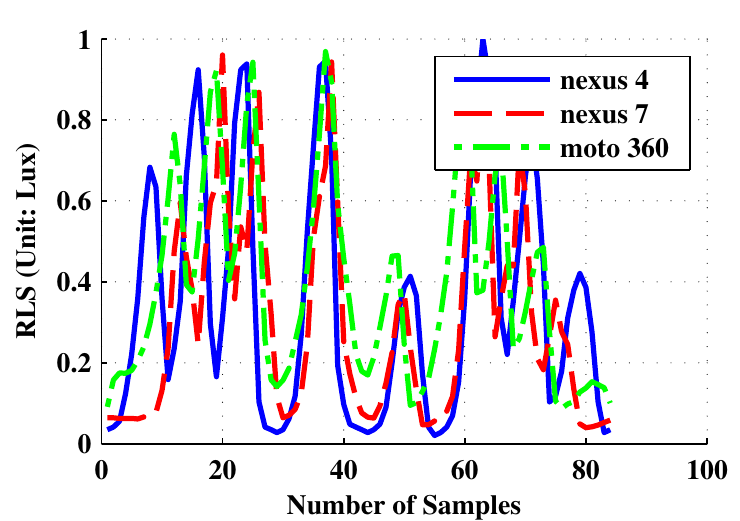}
    \label{fig:6:b}}
    \hfil
  \subfloat[Multiple light sources]{
    \includegraphics[width=0.3\textwidth]{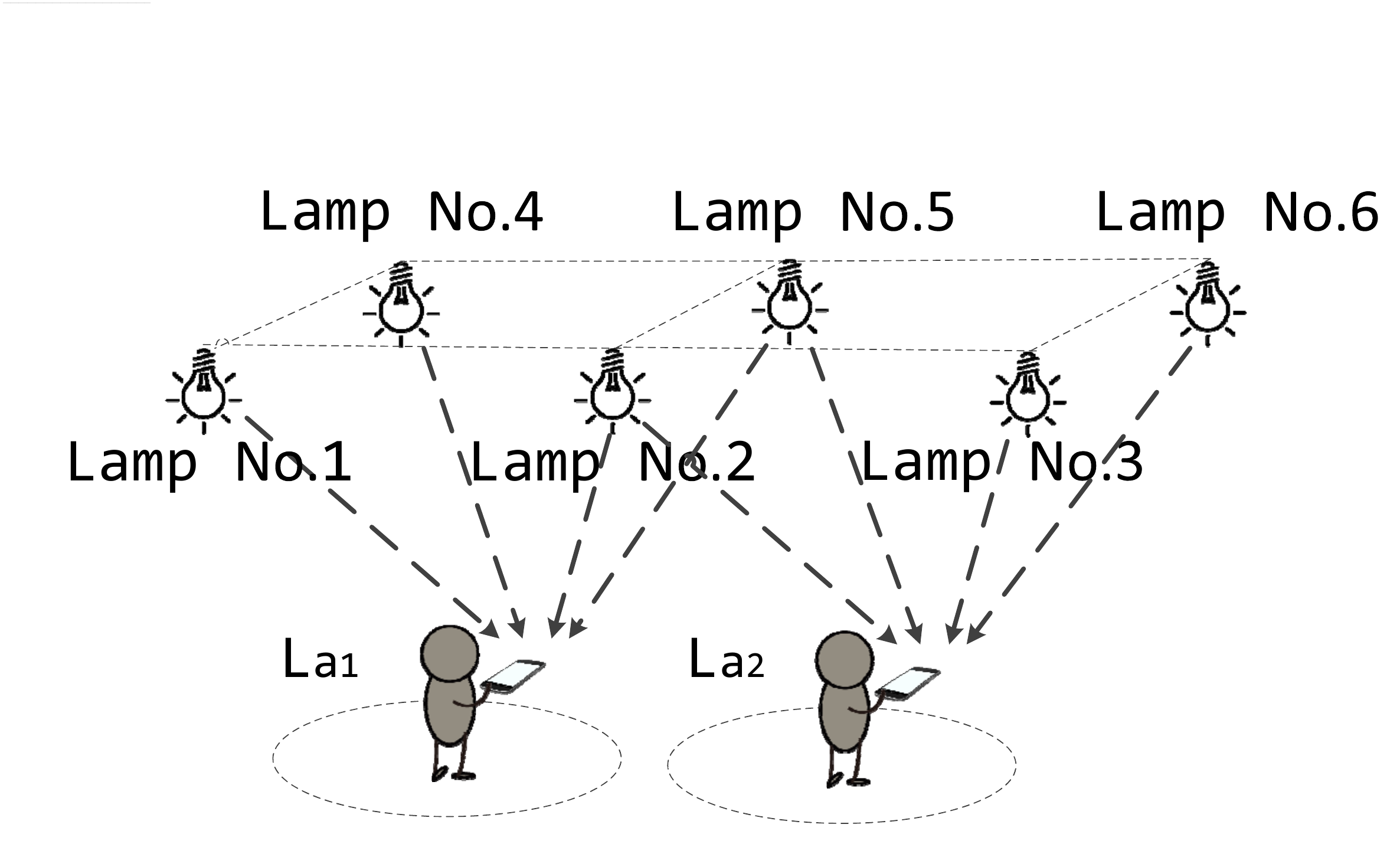}
    \label{fig:6:c}}
    \hfil
  \subfloat[Shading effect]{
    \includegraphics[width=0.22\textwidth]{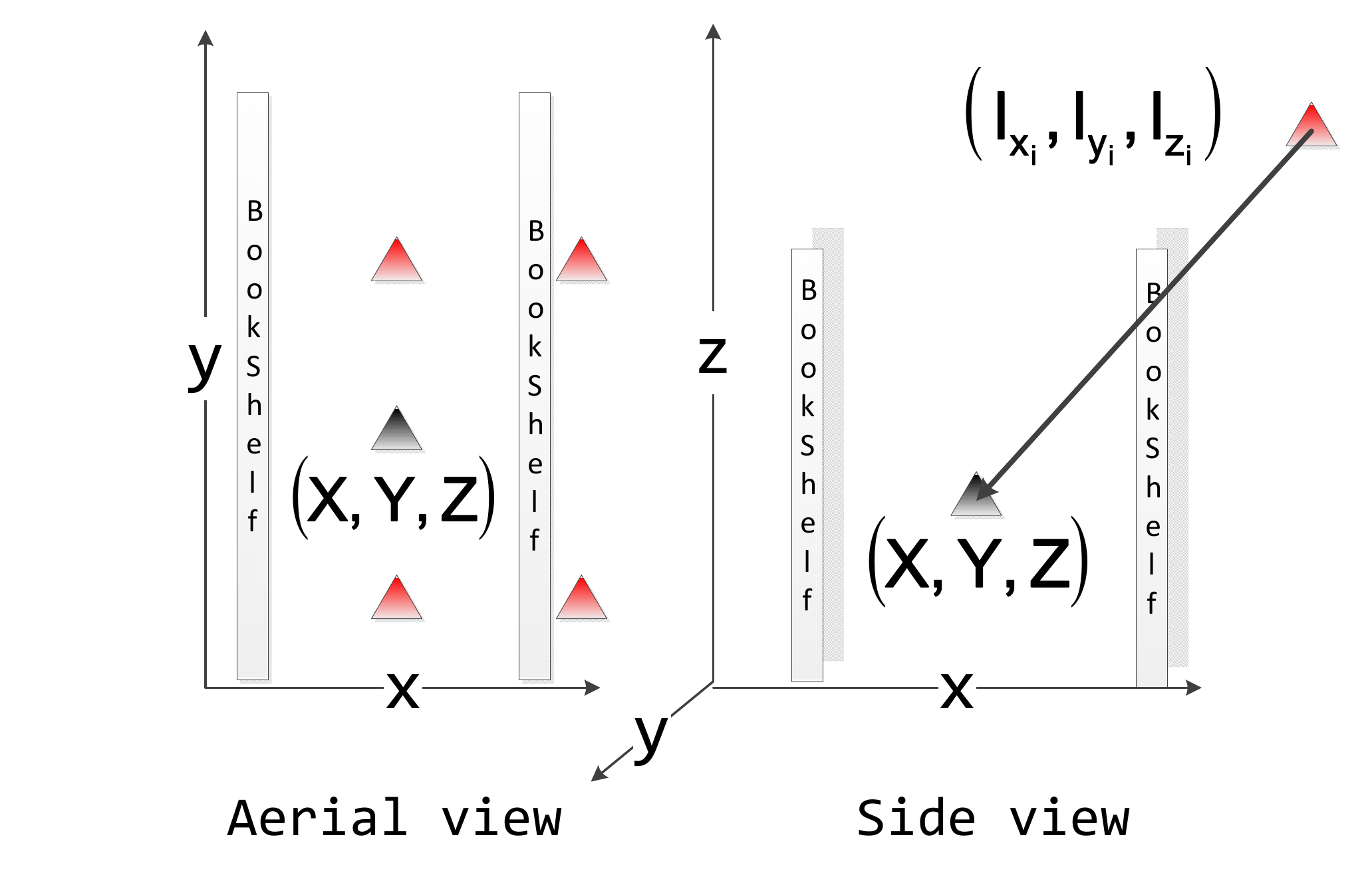}
    \label{fig:6:d}}
  \caption{(a) and (b): RLS trend captured by different devices. Despite the absolute value varies, Their trends$/$patterns are similar by normalizing and stretching.
  (c): multiple light sources.
  (d): shading effect of fixed obstacles. }
  \label{fig:6}
\end{figure*}

\textbf{Device Diversity:} A nature question is, \textit{does the basic model works well in describing the relationship between diverse devices and light sources?}
The answer is negative, due to the gain of diverse light sensors integrated in receiving devices makes RLS different.
We let one volunteer walk along a path with a constant speed, using a Google Nexus $4$, a Google Nexus $7$ and a smart watch Moto $360$ respectively to collect RLS.
Corresponding RLS trends are shown in Fig~\ref{fig:6:a}.
Fortunately, although the absolute RLS value captured by different devices varies, by normalizing and stretching, their RLS trends are similar, as shown in Fig~\ref{fig:6:b}.

To align the gain of different devices, we assume each device has a unique parameter $C$ that indicates its receiving gain.
As a result, the right-hand side of equation~\ref{eqn:rlsall} should pre-multiply $C$, to represent the relation between specified device and a target lamp.
Because both $C$ and $L_0$ in equation~\ref{eqn:rlsall} are constants, a once-for-good calibration is enough to calibrate each device.
For the sake of clear presentation, we leave out $C$ and use $L_0$ to represent the product of $C$ and $L_0$.

\textbf{Multiple light sources:} In scenarios like office, library and so forth, lamps are dense, and RLS is influenced by multiple lamps with a high possibility, as shown in Fig~\ref{fig:6:c}.
\textit{As a consequence, considering the geometric relationship between a single light source and receiving device is insufficient to calculate RLS precisely.}
Hence, we generalize the basic model to make it compatible with multiple light sources.
Assuming that there are $m$ lamps in the scenario, according to their positions, the user obtains the coordinates of $n$ closest lamps according to one's current position.
These lamps influence collected RLS obviously.
Assuming the contribution of these light sources are linear accumulated, we have:

\begin{equation}
L_a = \sum_{i=1}^n{L_{s_i}}
\label{eqn:multisources}
\end{equation}

Here $L_a$ is the RLS sampled by the light sensor, and $L_{s_i}$ indicates lamp $i$'s impact on the light sensor.
Basing on equation~\ref{eqn:rlsall}, we rewrite equation~\ref{eqn:multisources} to $L_a = \sum_{i=1}^n{L_{0_i}} \cdot p_i$, in which $p_i = \frac{\cos(\phi_i) \cdot e^{-(\frac{\theta_i - \theta_0}{\delta})^2}}{d_i ^ 2}$.
$\phi_i, \theta_i$ and $d_i$ are radiation angle, incidence angle and relative distance respectively, referring to lamp $i$.
In this equation, only $L_{0_i}$ is unknown, which indicates lamp $i$'s unique light strength characteristic.
So we could calculate the unique strength parameters of different lamps by solving following overdetermined equations in equation~\ref{eqn:mse}.

\begin{equation}
\begin{bmatrix}
L_{a_1} \\
... \\
L_{a_r} \\
\end{bmatrix}
=
\begin{bmatrix}
p_1 & p_2 & 0 & p_4 & p_5 & ...\\
... & ... & ... & ... & ... & ...\\
p_i & ... & p_j & ... & p_k & ... \\
\end{bmatrix}
\begin{bmatrix}
L_{0_1} \\
... \\
L_{0_m} \\
\end{bmatrix}
\label{eqn:mse}
\end{equation}

With the unique RLS parameter $L_0$ of each lamp, we can rebuild the light strength distribution by equation~\ref{eqn:phi}~$\sim$~\ref{eqn:multisources}.
Hence, each item in the six-dimensional variable space (roll, pitch, yaw and $3$D coordinate of the receiving device) has a corresponding RLS.

\textbf{Shading of Obstacles:} Obstacles such as bookshelves in the scenario may block the Line-of-sight (LOS) paths from the lamps to the light sensor, which makes RLS calculated by equation~\ref{eqn:multisources} deviated.
Basing on this observation, we leverage the geometrical relationship between receiving device, obstacles and lamps to model the shading effect.
Using the bookshelves as an example, we assume the user's current position is $(X_i, Y_i, Z_i)$, and the $n$ closest lamps' coordinates are $(l_{x_i}, l_{y_i}, l_{z_i}), i = 1, 2, ..., n$, as shown in Fig~\ref{fig:6:d}.
If $l_{x_i} = X_i$, the LOS path from the lamp to the light sensor exists.
If $l_{x_i} \neq X_i$, we connect $(X_i, Y_i, Z_i)$ and $(l_{x_i}, l_{y_i}, l_{z_i})$, to decide whether this path intersects with the bookshelf's framework between them.
We regard the path as a non-LOS path when they intersect, and eliminate this lamp's impact on RLS in equation~\ref{eqn:multisources}.

Hence, the extended model can be used to calculate RLS in arbitrary condition of the receiving device, which is validated in Section~\ref{sec:experiment}.
However, due to RLS may correspond to multiple positions (as shown in Fig~\ref{fig:2}), it is insufficient to uniquely locate user only by collecting RLS stationarily.
In the next section, we harness user's mobility to capture a spatially related RLS set, and then incorporate it and positions of existing lamps into a particle filter for positioning.

\section{Localizing Scheme using Particle Filter}
\label{sec:localizingmodule}
In this section, we illustrate the mechanism of our localizing scheme.
As depicted in Fig~\ref{fig:2:b}, a specified RLS value can be captured at multiple locations.
Intuitively, we try to exploit the temporal and spatial relationship between consecutive RLS, in order to eliminate the position ambiguity caused by a single RLS.
Particle filter is a set of on-line posterior density estimation algorithms that estimate the posterior density of the state-space by directly implementing the Bayesian recursion equations, which is exactly the case of Lightitude.
As a result, we adopt the concept of particle filter to design our localizing scheme.

The key idea underlying our scheme is that, harnessing user's mobility, we exploit a set of RLS captured by the user to eliminate impossible candidate positions, whose RLS deviates significantly from that captured by the user.
A biased stride length and heading generate candidate traces, whose corresponding RLS set is unmatched with what the user has collected.
\textit{As a result, user's position, together with stride length and heading converge simultaneously after enough iterations.}
We divide the basic localizing scheme into four parts: \textbf{initialization stage, attributing weights to particles, continuous walking} and \textbf{achieving convergence}.

However, this basic positioning scheme is easily influenced by user's behaviours and complex environmental factors.
\textbf{Sunlight interference} greatly influences the indoor light strength distribution, and its light strength is almost an order of magnitude greater than that of light sources.
What's worse, the light sensors integrated in most commodity smart devices can't discern the difference between sunlight and light emitted from light sources.
At the meantime, encountering \textbf{unpredictable behaviours} of users like picking a phone call or putting the device in pocket will mislead Lightitude, due to RLS deviates severely from what it should be in these conditions.
Furthermore, \textbf{shading effect of human-body} has a risk to deviate collected RLS also.
Facing all these dilemmas, we design mechanisms to cope with them separately, and finally make Lightitude can locate user against various kinds of interference.
Besides, we \textbf{coexist the localizing scheme with WiFi} to achieve a higher accuracy and a faster convergence speed.

\subsection{Basic localizing scheme}
\label{sec:basiclocalizingmodule}
\textbf{Initialization: } At the initialization stage, we scatter particles uniformly in the scenario with granularity $G$ (e.g., one particle in $0.1 \times 0.1 \times 0.1 m ^ 3$).
Each particle is possible to be user's current position.
Combining with the floor plan, positions of the lamps and current rotation matrix captured by the device, RLS of these particles are calculated by the light strength model.
Comparing with RLS captured by user, the particle with a closer RLS has a bigger weight, and vice versa.
Specially, the particle with the strongest weight represents the most possible position of the user.
After obtaining the weight of all particles, only part of particles $N$ with strong weight are chosen as the initial particles of the next stage, in order to reduce the computation overhead.

\begin{figure*}[t]
  \centering
  \subfloat[Band for ``children'' particles]{
    \includegraphics[width=0.28\textwidth]{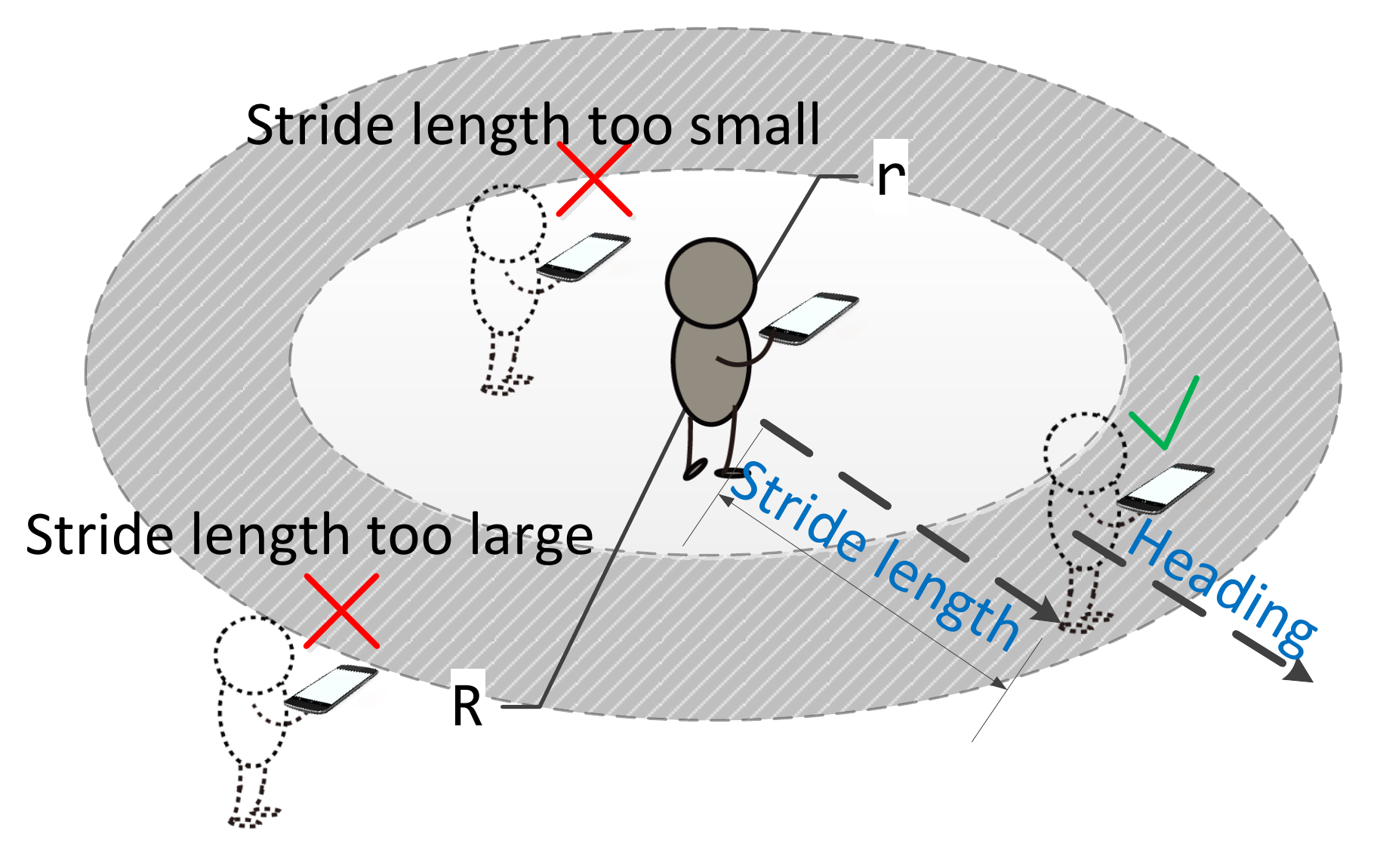}
    \label{fig:8:a}}
    \hfil
  \subfloat[Interference of Sunlight]{
    \includegraphics[width=0.25\textwidth]{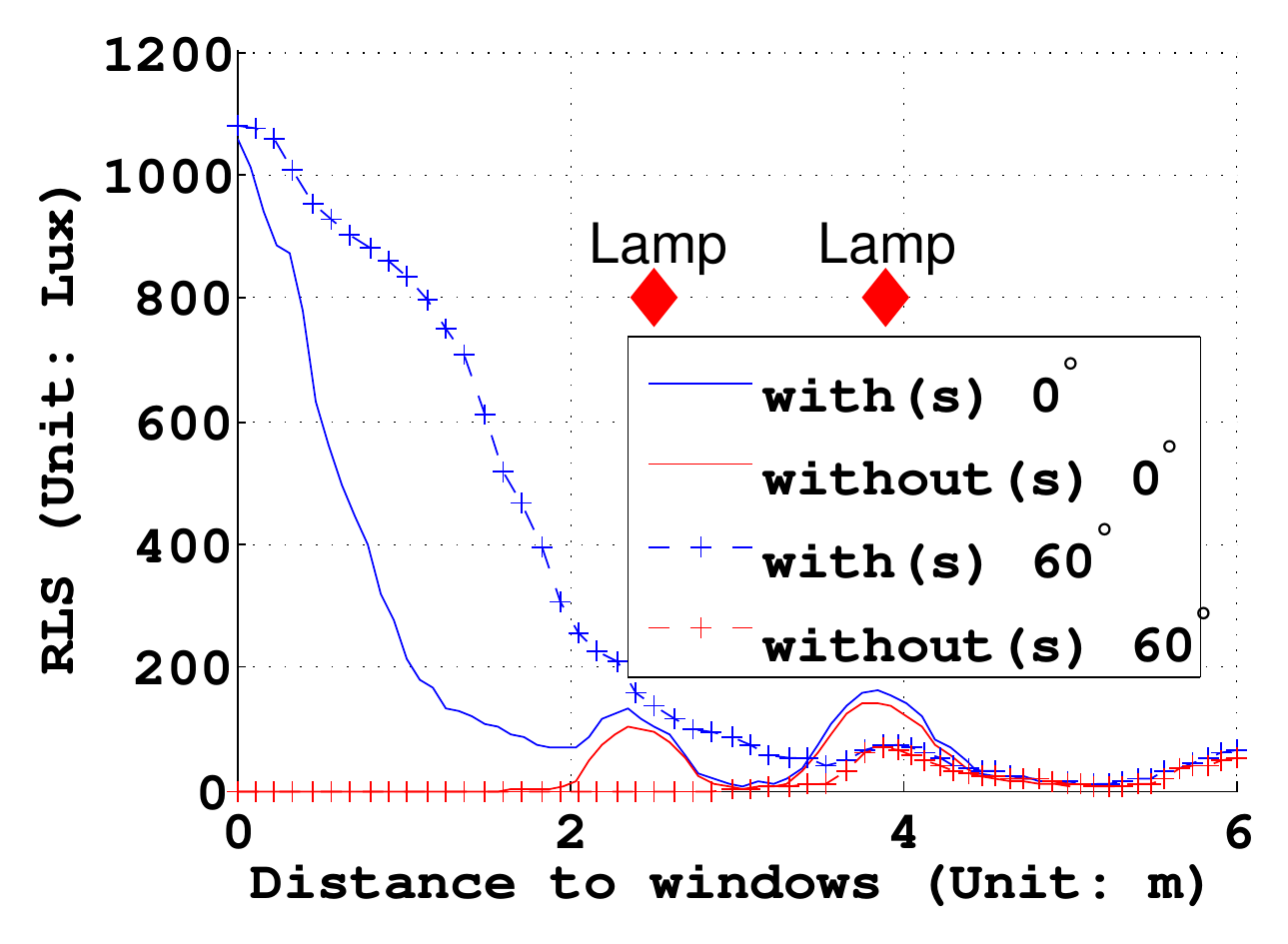}
    \label{fig:8:b}}
    \hfil
  \subfloat[Office overview]{
    \includegraphics[width=0.23\textwidth]{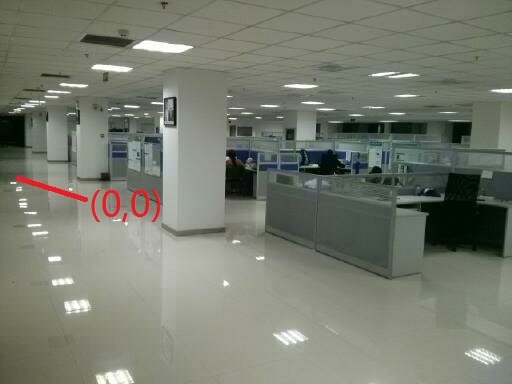}
    \label{fig:8:c}}
    \hfil
  \subfloat[Library overview]{
    \includegraphics[width=0.13\textwidth]{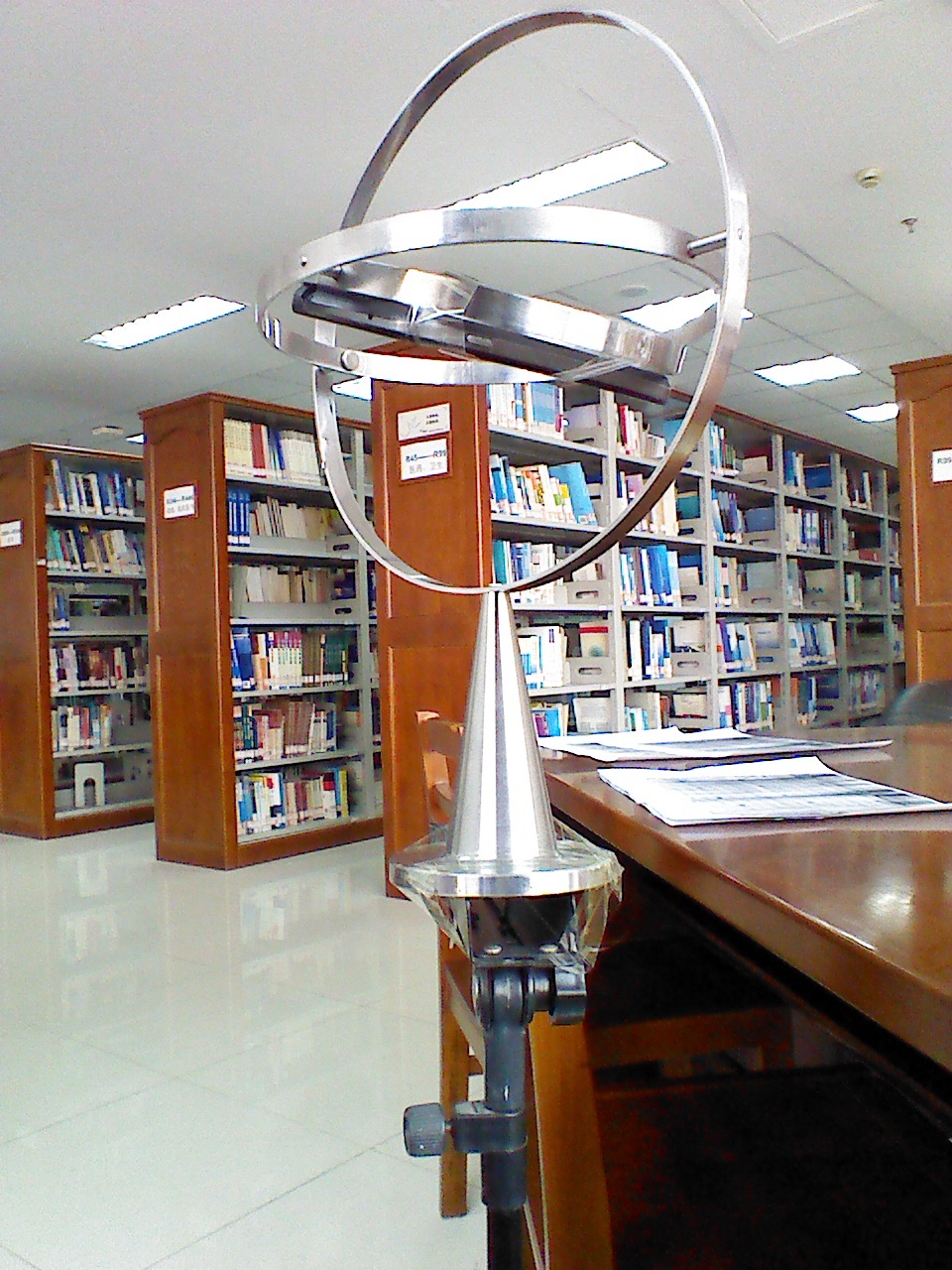}
    \label{fig:8:d}}
    \hfil
  \caption{(a): Exploiting the relation between successive particles to generate candidate particles. Feasible stride lengths and headings are generated simultaneously.
  (b): Interference of sunlight. RLS with sunlight are collected at $12$ a.m. in a fine day, while RLS without sunlight are collected at $9$ p.m. at the same day.
  (c) and (d): Experiment scenarios.
  (d) also shows the gyroscope used for data collection.}
  \label{fig:8}
\end{figure*}

\textbf{Attributing weights to particles:} Surprisingly, a particle is associated with multiple RLS in our localizing scheme.
This fact seems conflict to the light strength model: an item in the six-dimensional space, in other words, a position/particle with specified rotation state corresponds to a single RLS.
The reason for this conflict is that, we run the localizing module once detecting a step of the user.
Since running the localizing module once upon receiving a modified RLS seems more reasonable, possessing numerous particles within sub-seconds causes a great computation overhead. (Average $10$ RLS samples can be captured in a single step, near the projection of a light source.)
Meanwhile, light sensors work in on-sensor-changed rule, i.e., sample once detecting a modified RLS.
As a consequence, the sample time of RLS is not regular.
To conclude, it's hard to gauge the gap between adjacent particles, thus makes positioning accurately impossible.
Hence, we adhere a RLS set, other than a single RLS with each particle.

\begin{figure}[b]
  \centering
  \includegraphics[width=0.4\textwidth]{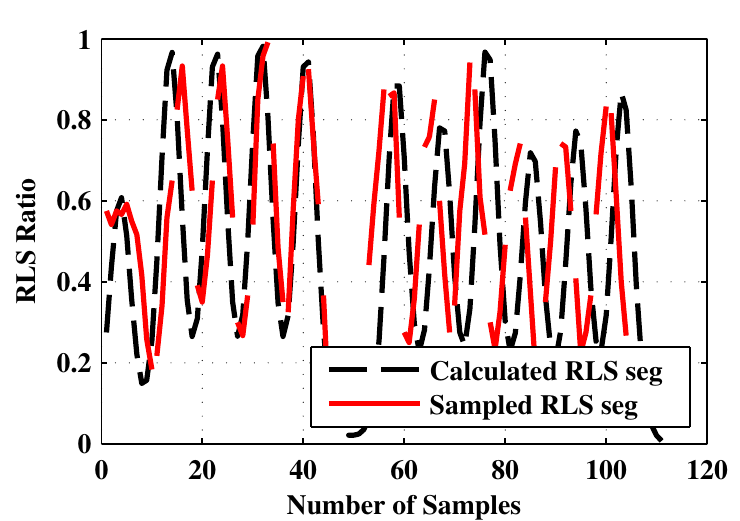}\\
  \caption{Despite the RLS trend is similar, shift will makes L2 norm distance (Euclid distance) fails in calculate the weight of the particle. }
  \label{fig:7}
\end{figure}

In a single step of the user, one collects multiple RLS with a big possibility.
We align the timestamps of RLS with that of corresponding inertial data (step), to fetch the corresponding RLS set in a single step.
However, simply picking a single of them to present current particle can not take full advantage of collected data.
Thus, instead of using only a single RLS, we use all of them for positioning.
Fig~\ref{fig:7} shows a RLS trend collected by walking $30$ steps and the corresponding one calculated by the light strength model.
The applicability of the calculation is that, we understand the particle's $3D$ position and rotation matrix.
Leveraging the light strength model, we can calculate a RLS set within an arbitrary stride length, with a specified granularity.
As a result, we can get the RLS set of all particles within a single step.

Similarly, comparing with the ground truth RLS set, the particle with a close RLS set obtains a bigger weight.
An intuitive idea is to calculate the Euclid distance between these two sets.
However, as mentioned before, RLS is not uniquely captured in a single step: the movement of the user is not regulatory and the the rule of light sensors is on-sensor-changed.
As a consequence, these influence factors cause shifts and unequal length in the two trends, as shown in Fig~\ref{fig:7}.
So simply adopting Euclid distance calculation is not appropriate.
By observing the characteristics of these two trends, we employ dynamic time warping (DTW)~\cite{keogh_exact_2005} as the distance measure to compute the distance between two RLS trends.
Since DTW is an elastic measure, it can handle the misalignment in two trajectories.
Using the collected RLS trend as an input, we calculate current weight of each particle $w$ by equation~\ref{eqn:particleweight}.

\begin{equation}
w = \frac{1}{e^{\frac{DTW( L_c, L_m )}{K}}}
\label{eqn:particleweight}
\end{equation}

In this equation, $L_c$ is the RLS trend collected by the light sensor, $L_m$ is the corresponding calculated one.
In ideal condition, $L_c = L_m$, so the weight of the particle at ground truth position equals $1$.
$K$ is a smoothing factor, and a bigger $K$ makes weight of particles more smooth that decreases the convergence speed.
Possessing equation~\ref{eqn:particleweight}, we give each particle a corresponding weight.

\textbf{Continuous walking: }After initialization stage, the user walks to obtain spatially related RLS.
We take the starting particles as ``mother'' particles and their follow-ups as ``children'' particles.
They are separated by a single stride length of the user.
We assume the user's minimum stride length is $r$ and maximum stride length is $R$, consequently a ring with radius $S$ centering at the ``mother'' particles, ranging from $r$ to $R$ is the band for their ``children'' particles, as shown in Fig~\ref{fig:8:a}.
Particles that fall out of this range are eliminated, as well as the ones that violate the floor plan.
Otherwise they are preserved as candidate particles.
Similar to the initialization stage, each candidate particle obtains a weight that derives from its RLS.

To exploit the relation between consecutive particles, we multiply weight of each particle with weight of its ``mother''  particle,
in order to draw a trajectory which indicates the user's ground truth trace.
Given the efficiency problem, we only choose a slice of candidate particles with strong weight as the starting points in the next iteration.
As a result, particles with small weights are eliminated before hand.
Survived particles with same $3D$ coordinates are regrouped by summing their weights together.
Finally, their weights are normalized and sorted at the end of each iteration.
These particles are regarded as the input of next iteration of the localizing scheme.

\textit{Besides user's position, our localizing scheme also gets user's stride length and heading.}
Orientations from ``mother'' particles to ``children'' particles are recorded as user's candidate headings, and their gaps are recorded as user's candidate stride lengths.
If the user changes one's heading in a big scale (e.g., encounters a corner), a sudden change is detected by the gyroscope integrated in the device.
If no sudden change is detected, Lightitude assumes the user doesn't change one's heading, and newly generated particles shall inherit the orientations of their ``mother'' particles.
Otherwise, Lightitude discards the inheritance rule; newly generated particles are scattered uniformly in the rings around their 'mother' particles.

\textbf{Achieving convergence: }In most cases the particles converge after a few steps of the user.
Specifically, the convergence is achieved whenever one of the two cases happens: 1) all particles are in a small area (e.g., $0.5 \times 0.5 \times 0.1 m^3$);
2) one particle has an exceedingly powerful weight than other particles (e.g., bigger than $50\%$ of the overall weight).
However, in quite a few cases, the particles do not converge due to an ambiguous RLS trend.
In this case, after a fixed number of the user's steps, Lightitude terminates itself mandatorily and chooses the particle with the strongest weight as the ground truth position.

With the help of convergence conditions, the user can terminate the localizing module in a few steps.
However, in many conditions, the user walks longer than the distance needed for convergence.
Exploiting this long trace collected by the user, Lightitude achieves a more accurate positioning result.
We show the performance of Lightitude under this condition in Section~\ref{sec:experiment}.

\begin{figure*}[t]
  \centering
  \subfloat[Deviate RLS Ratio]{
    \includegraphics[width=0.23\textwidth]{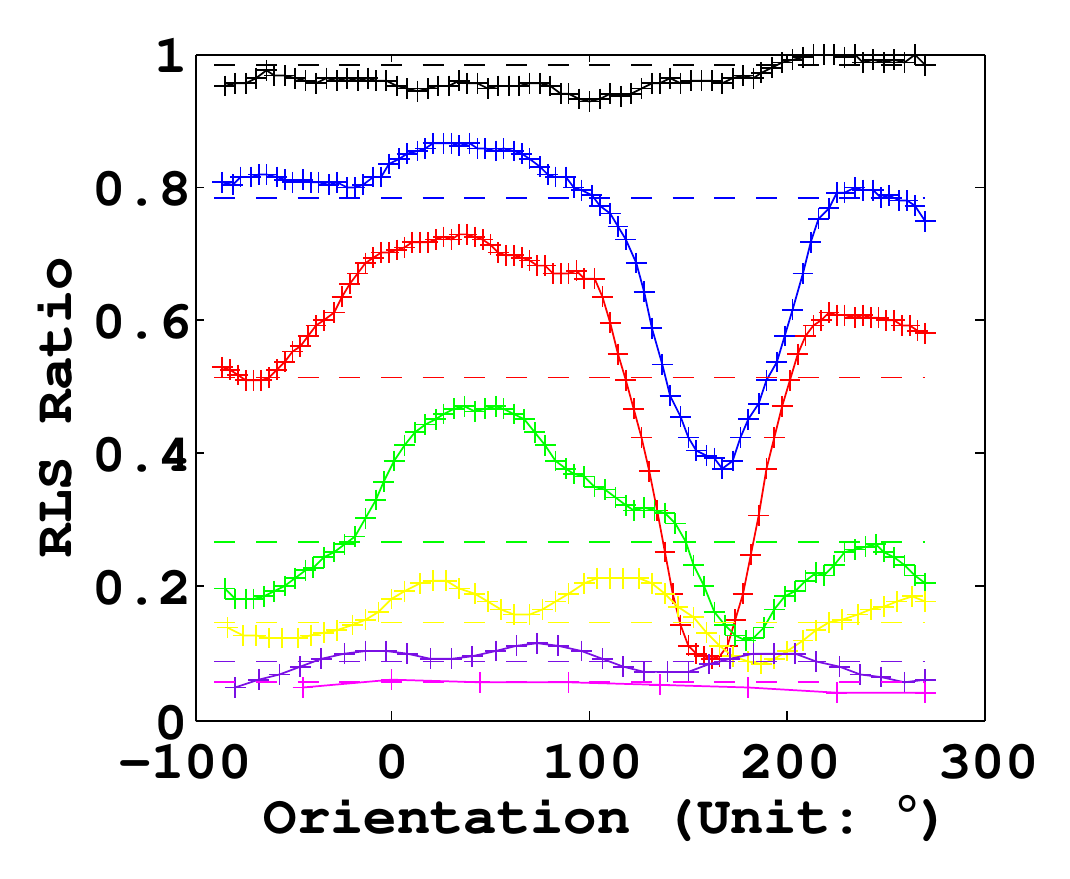}
    \label{fig:9:a}}
    \hfil
  \subfloat[Deviate Weight]{
    \includegraphics[width=0.23\textwidth]{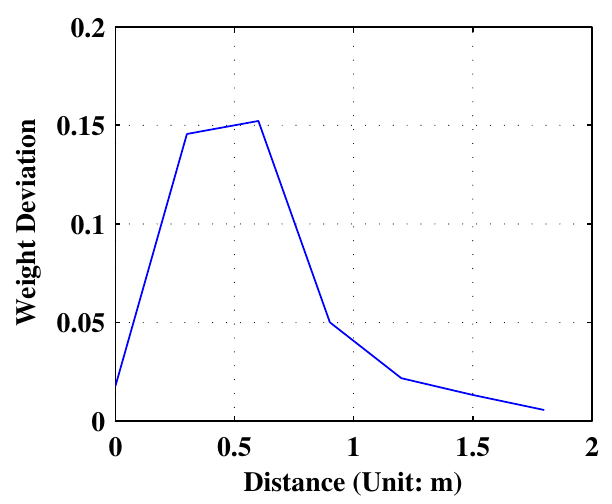}
    \label{fig:9:b}}
    \hfil
  \subfloat[Rotation status change]{
    \includegraphics[width=0.23\textwidth]{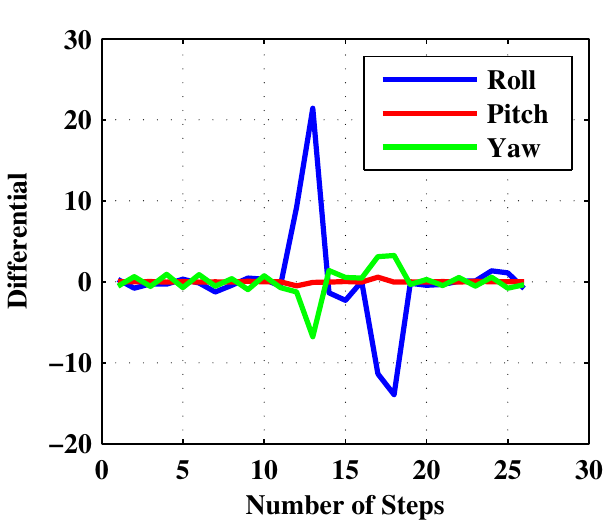}
    \label{fig:9:c}}
    \hfil
  \subfloat[RLS Change]{
    \includegraphics[width=0.23\textwidth]{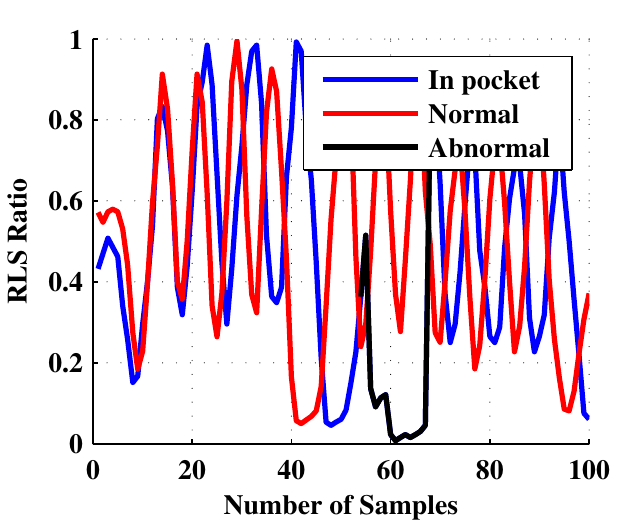}
    \label{fig:9:d}}
  \caption{(a) shows the human body's shading effect under different conditions. From top to bottom, each $+$ curve indicates RLS ratio w.r.t. its orientation at distance $0$m, $0.3$m, $0.6$m, $0.9$m, $1.2$m, $1.5$m, $1.8$m. Corresponding $-$ curve is the ideal RLS $L_0$ calculated by the light strength model.
  (b) shows the mean weight deviation in seven different conditions.
  (c) shows the status change of the device when detecting user's behaviour.
  (d) shows shows the RLS change of the device when detecting user's behaviour. Two traces have a little shift due to volunteer-diversity.}
  \label{fig:9}
\end{figure*}

\subsection{Localizing against interference}
\label{localizingagainstinterference}
Upon using the localizing scheme, some unpredictable conditions may be encountered, such as interference of sunlight, interference of human-body and unpredictable behaviours of the user.
We design mechanisms to cope with these conditions separately.

\textbf{Interference of sunlight: } We find that sunlight interference is not only a challenge, but also an opportunity to boost Lightitude's performance.
We first discuss the challenge and the opportunity separately, and then propose a mechanism that enhance the robustness of our original localization model, by leveraging the advantage of sunlight that we observe.

The challenge is that, \textit{the light sensors integrated in most commodity smart devices can't discern the difference between sunlight and light emitted from light sources.
Moreover, the light strength of sunlight is almost an order of magnitude greater than that of light sources, which greatly influences the indoor light strength distribution.}
As shown in Fig~\ref{fig:8:b}, by keeping the receiving device flat at fixed height $1$m in the office scenario, RLS is influenced by sunlight within $2.5$m away from window;
keeping the receiving device $60^\circ$ deviated from the horizontal plane (tilt to the window), the distance is $3.9$m.
Meanwhile, quantifying sunlight's impact is a formidable task because it varies at different times of the day and in different weather conditions.

The opportunity is that, \textit{the positions with interference of sunlight are restricted in limited areas in most indoor environment.}
Accordingly, an RLS that influenced by sunlight provides us a hint to scatter particles in sunlight-coverage areas, rather than scattering them uniformly in the scenario.
This hint reduces Lightitude's search scope and gives a boost to its performance.

Though sunlight has some impact on our original localization model, we enhance its robustness by leveraging the advantage of sunlight that we observe.
Using equation~\ref{eqn:phi}~$\sim$~\ref{eqn:multisources}, we rebuild the light strength distribution.
Furthermore, we check if a specified RLS exists in this distribution and infer its corresponding orientation and altitude.
If a user collects an RLS trend which contains a very strong segment that cannot be in the distribution, or with an infeasible device's status (e.g., height = $2.5$m), we regard this segment as influenced by sunlight.
In this case, we modify the original particle-scattering rule in the raw localizing module.
Instead of scattering them uniformly in the scenario, we scatter all particles in the sunlight interference area (e.g., two regions with $2.5 \times 30m^2$ near the windows in our library scenario).
We determine the areas with sunlight's interference in advance by the work in Fig~\ref{fig:8:b}.
All these particles preserve only IMU information for back-propagation, until its ``offspring'' particles break out the sunlight interference area.
Original particle-scattering rule is adopted again afterward.
Experiment result validates our claim, as shown in Section~\ref{sec:experiment}.

\textbf{Influence of Human-body's Shading: } To test human-body's shading, we conduct experiments in different scenarios, which takes account of device's orientation, altitude, distance to projection of the nearby lamp, and its relative position with user's body. The following experiments show that, despite human-body's shading effect weakens RLS, it has little impact on Lightitude's performance.

A user's body will block several LOS paths from the light sources to the device, which further changes its corresponding RLS.
One obvious observation is that, if a user's body blocks a LOS path from a strong light source to the receiving device (e.g., the $0.3$m case with $180^\circ$ in Fig~\ref{fig:9:a}), the particle at current ground truth position has a big weight loss, for $L_c$ deviates from $L_m$ heavily.
On the other hand, if a user walks in a relative dark area, $L_m$ is weak, and user-body's shading has little impact on its value (e.g., the $1.8$m case in Fig~\ref{fig:9:a}).
As a result, its weight will not have a big deviation.
Hence, we focus on the condition that $L_c$ heavily deviates from $L_m$.

To get the strongest RLS deviation, we first investigate in which status can the device receive the strongest RLS.
As mentioned in Section~\ref{sec:lightModel}, a high altitude and a small incidence / radiation angle of the receiving device help increasing RLS.
In user's route, device's altitude is relatively stable. On the other hand, the incidence / radiation angle changes frequently, according to different lamp sets.
Basing on this observation, keeping the device's height fixed at $1$m in the library scenario, we have light strength contours under different orientations of the device (yaw $0^\circ$, $30^\circ$, $60^\circ$).
We find that comparing with other orientations, keeping the device flat ($0^\circ$) captures a maximum number of strong RLS. As an example, setting $20$ Lux as the threshold of strong RLS, $31.87\%$, $27.64\%$, $12.79\%$ of the calculated RLS exceed the threshold in these three scenarios respectively.
So we select the device-flat scenario as the experiment scenario to explore human-body's shading effect.

We conduct experiments of human-body's shading in different conditions.
One volunteer is recruited to hold the receiving device with its altitude fixed at $1$m, and keep the device adhering to one's body, by which human-body's shading influences RLS most.
The volunteer selects $10$ arbitrary lamps, and changes the distance between the projections of the device and the lamp: 1) $0$m (standing right under the lamp); 2) $0.3$m; 3) $0.6$m; 4) $0.9$m; 5) $1.2$m; 6) $1.5$m; 7) $1.8$m.
In each condition, the volunteer moves one's body around the device to get the shading effect.
As a result, collected RLS ratio is shown in Fig~\ref{fig:9:a}.
We notice that shading effect of human-body has a big impact on RLS occasionally, and the impact peaks with one's back to the lamp (around $180^\circ$).
Corresponding weight deviations of these conditions are shown in Fig~\ref{fig:9:b}.

Despite the shading effect of human-body influences RLS, it has little impact on Lightitude's performance.
The key insight is that, the shading effect occurs occasionally, and can be corrected by using the adjacent samples which are less affected by shading.
As shown in Fig~\ref{fig:9:b}, weight deviation occurs mainly in a single step of the user (in other words, a single iteration).
When a user walks across a lamp's coverage area, Lightitude operates for several iterations.
Biased weights mainly affect a single iteration, while other iterations are unaffected.
Furthermore, Lightitude is robust against the weight deviation of a single iteration.
The reason is that, in the localizing module, \textit{the weight of each particle multiplies the weight of its ``mother'' particle, and this dilutes its biased weight by the superior performance of its ``mother'' / ``children'' particles.}
At last, as mentioned in Section~\ref{sec:localizingmodule}, the weight of candidate particles are normalized at the end of each iteration, accompanying with other hundreds of particles. This dilutes the biased weight of a single particle additionally.

To conclude, the shading effect of human-body influences RLS occasionally, but it has little impact on Lightitude's performance.
In section~\ref{sec:experiment}, we conduct experiments under different device-holding gestures to validate Lightitude's robustness.

\textbf{Unpredictable behaviours of the user: } The normal localization module has a risk to be misled by user's unpredictable activities.
In using Lightitude, a phone call will make the user put the phone in the vicinity of one's face, by which RLS degrades immediately in an unpredictable way.
In another case, the user puts the receiving device in one's pocket, by which RLS will remain low (near $0$).
\textit{These activities mislead Lightitude, since such positions with unexpected RLS does not exist in the RLS distribution of the scenario.}
Facing these dilemmas, we design a prevention module for Lightitude, in order to increase the robustness of Lightitude even facing with pre-mentioned unpredictable behaviours.

We recruit one volunteer with average stride length $0.8$m to walk in two different behaviours.
In normal case, one puts the receiving device in an arbitrary status, then keeps it quite stable and walk along the path.
In another case, one puts the receiving device quite stable, walk along the path with an arbitrary status also, but put the device in the pocket after walking $12$ steps, and put it out again after walking another $4$ steps.
We name this $4$ steps as the abnormal trace.
As depicted in Fig~\ref{fig:9:c}, at the start of the abnormal trace, together with RLS, the roll/pitch of the device changes obviously.
At the end of the abnormal trace, RLS and roll/pitch of the device changes obviously again.
We use these two trademarks to indicate the start and the end of the user's behaviour.
In the abnormal trace, we pause the localizing module; only pedometer works to record the step number, in order to count the length of the abnormal trace.
After recovering from the abnormal trace, all candidate positions add the distance of the abnormal trace, and the normal localizing module starts again.

By integrating this prevention module, Lightitude can tolerate some unpredictable behaviours of the user.
However, despite processing this mechanism, there always exist behaviours which are out of Lightitude's consideration.
To avoid error-positioning, the most dependable solution is restarting Lightitude.
But in designing Lightitude, we take the user's cost as the first priority, and try to make Lightitude a once-for-good deal for users.
Under this principle, in future applications of Lightitude, more behaviour scenarios/patterns can be added into this prevention module to enhance the robustness of Lightitude.

\subsection{Coexisting With WiFi}
\label{sec:coexistingwifi}
Light-based positioning requires no pre-deployed infrastructures.
But if multiple similar paths exist, in other words, the light sources in these paths share similar lighting characteristics, the user need to walk a longer distances for convergence.
Facing this dilemma, in this section, we integrate a WiFi fingerprinting module with the light-based positioning module, in order to design a fine-grained and fast-convergence positioning scheme.
We first borrow the design of RADAR~\cite{bahl2000radar} to design a WiFi fingerprint-based module.
Then we use it to provide a candidate position set as the start points of our localizing scheme,
further exploit our localizing module to prune this set.
We expand the details of our method below.

\textbf{Generating candidate set: }Different from WiFi fingerprint-based schemes like RADAR who directly provide a coordinate as the positioning result, our fingerprint-based schemes provide only a \textit{candidate set}.
The candidate position set contains the top $M$ most possible positions of the users, not necessary the only one with the biggest possibility.
By using which we can refine the initialization stage of our localizing scheme: \textit{particles with close RLS set but with distant fingerprint are eliminate beforehand.}
However, how to determine the size of the candidate set is a puzzle.
Candidate set with a big size increases the computation overhead, and candidate set with a small size has a huge risk to miss the ground truth position, due to fingerprint mismatch.
We expand our methodology and implementation on searching for the most appropriate size of candidate set below.

For a specific area in which fixed numbers of APs has been deployed, we choose a candidate set with proper size according to the area size, in order to ensure the ground truth position of the user is in it.
In addition, the size of candidate set is not only associated with the area of the target scenario, but also will be influenced by the fingerprint numbers in this area.
Taking these two influence factors together, we formulate the relationship between the size of the candidate set, area size and fingerprint granularity.

\begin{figure*}[t]
  \centering
  \subfloat[Cross-section (Office)]{
    \includegraphics[width=0.21\textwidth]{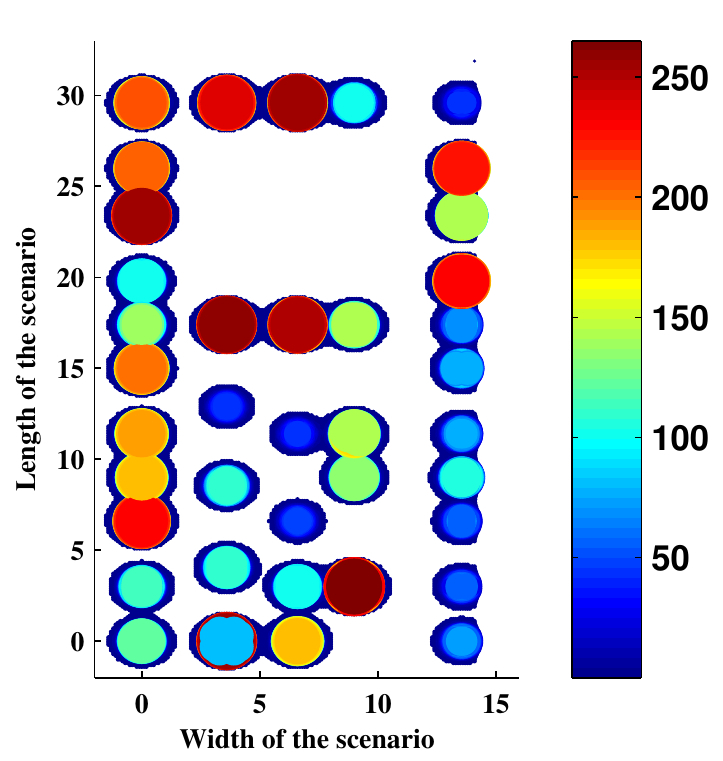}
    \label{fig:11:a}}
    \hfil
  \subfloat[Cross-section (Library)]{
    \includegraphics[width=0.32\textwidth]{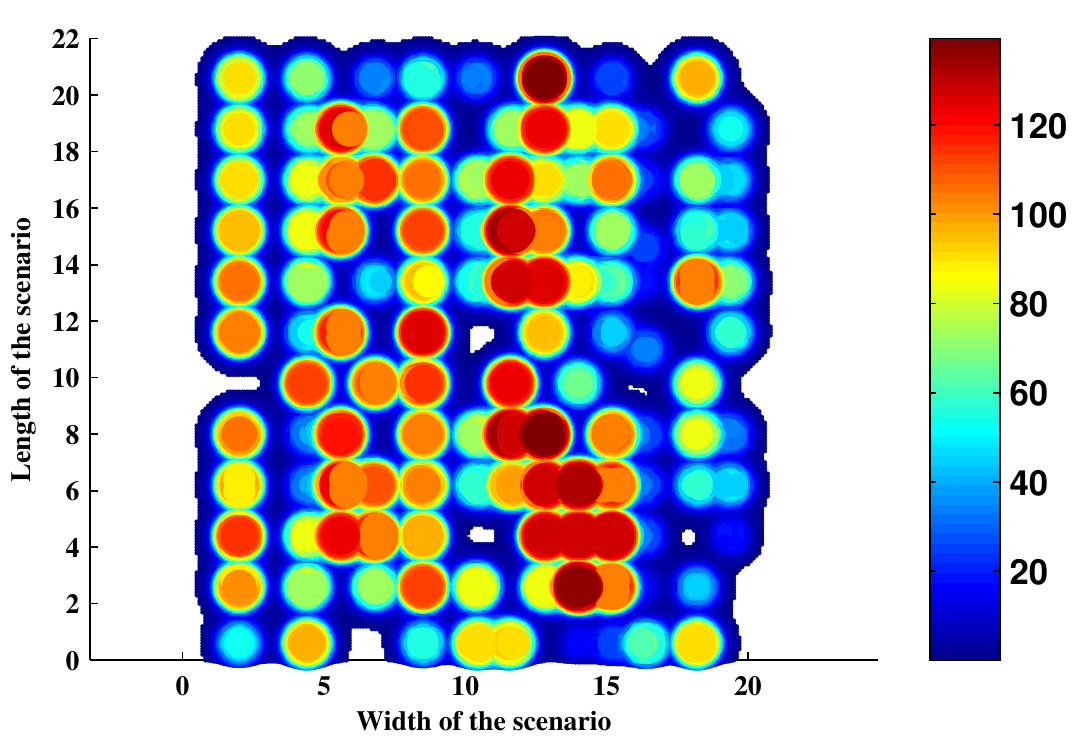}
    \label{fig:11:b}}
    \hfil
  \subfloat[Light strength model]{
    \includegraphics[width=0.21\textwidth]{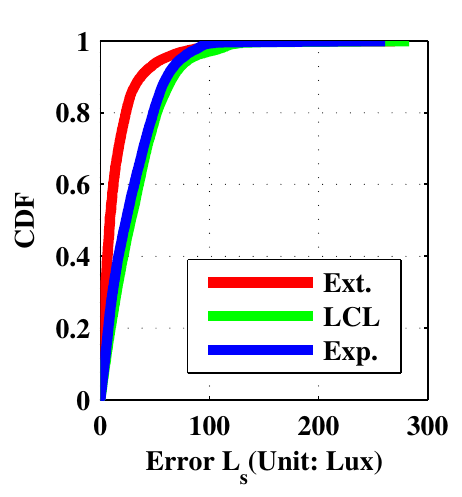}
    \label{fig:11:c}}
    \hfil
  \subfloat[RLS deviation in different $\theta$]{
    \includegraphics[width=0.21\textwidth]{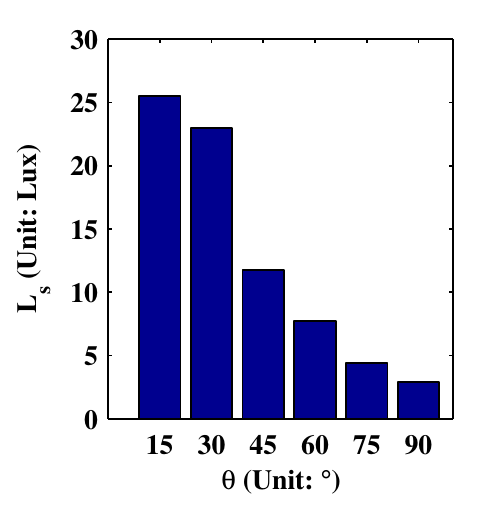}
    \label{fig:11:d}}
  \caption{(a) and (b) are cross-sections of the light strength distribution, when volunteers face the device upright and keep its altitude fixed at $1$m.
  (c) shows the performance of different light strength models. Ext. is short for extended model, Exp. is short for exponential rule (square).
  (d) shows the mean deviation in different $\theta$. In x axis, $15$ means $0^\circ \sim 15^\circ$, and so on. }
  \label{fig:11}
\end{figure*}

First, we assume the size of the experiment scenario $A$ is $m \times n$ (unit: $m^2$), and in which a single fingerprint lies uniformly in each area with $m_1 \times n_1$ (unit: $m^2$).
So in this scenario, the number of the fingerprint $F = \frac{m \times n}{m_1 \times n_1}$.
We randomly choose area $A'$ with size $m' \times n'$, so in this area, we have fingerprint number $F' = F \times \frac{m' \times n'}{m \times n}$.

In $A'$, the positioning accuracy is determined by the $F'$.
If only a single fingerprint lies in $A'$ ($F' = 1$), the user can precisely locate oneself by running the WiFi fingerprinting module, because user's ground truth position is right near the sole fingerprint.
We name this fingerprint as the \textit{ground truth fingerprint}.
If multiple fingerprints exist in $A'$, the ground truth fingerprint may not be the result of the WiFi fingerprinting module, due to fingerprint mismatch.
Suppose the ground truth fingerprint ranks $k$th in the top candidates of the result set, so in area $A'$, picking out a candidate set with size $k$ can ensure the ground truth fingerprint is in this set.
To conclude, the size of candidate set $M$ is determined by the fingerprint number $F'$ in $A'$.

\begin{figure}[b]
  \subfloat[$M$ and $m', n'$]{
    \includegraphics[width=0.23\textwidth]{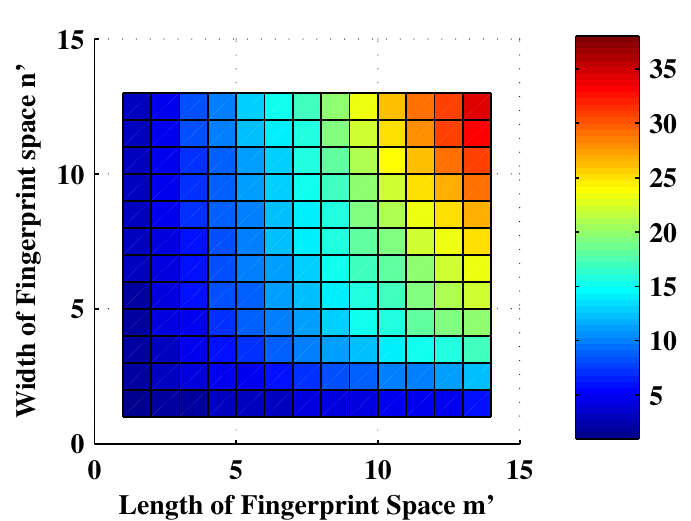}
    \label{fig:10:a}}
    \hfil
  \subfloat[$M$ and $F'$]{
    \includegraphics[width=0.23\textwidth]{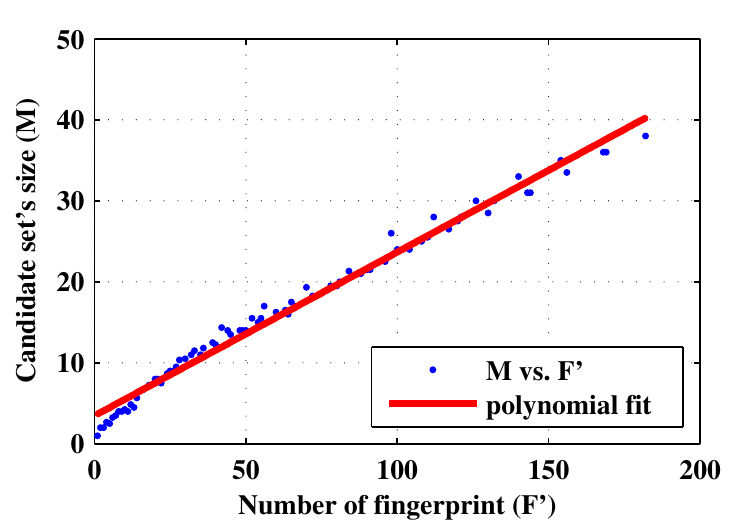}
    \label{fig:10:b}}
  \caption{(a): Relationship between $M$ vs. $m', n'$. For a target length/width of the fingerprint space, the size of candidate set $M$ is determined, as depicted by the color bar.
  (b): Relationship between $M$ vs. $F'$. We use a polynomial fit to describe the relationship between $M$ and the number of fingerprints $F'$.}
  \label{fig:10}
\end{figure}

To formulate the relationship between the $M$ and $F'$, we conduct extensive benchmarks in our experiment scenario.
We choose $182$ areas with different size and fingerprint numbers as the test set.
In each area, we have a fingerprint test set and its corresponding fingerprint database.
We run the WiFi fingerprint module on each fingerprint in the test set.
As a result, we get the candidate set of size $F$ that contains the ground truth fingerprint.
Overall, the relationship between $F$ and the size of fingerprint space $m', n'$ is depicted in Fig~\ref{fig:10:a}.
In addition, we get the relationship between $F$ and $N'$, as shown in Fig~\ref{fig:10:b}.

We notice in Fig~\ref{fig:10:b}, the relationship between $M$ and $F'$ is almost polynomial.
So we use polynomial function with degree $1$ to describe their relationship.
As a result, $M = p_1 \times F' + p_2$.
In our experiment scenario, $p_1$ and $p_2$ is $0.2$ and $3.5$ separately.
As a result, giving an area with a fixed fingerprinting granularity, we can directly get the size of candidate set $M$.
$M$ contains the ground truth fingerprint, which provides a robust input for further pruning, by using our light-based localizing module.

\textbf{Pruning candidate set: } We use our light-based localizing module to further prune the candidate set, in order to locate the user preciously and quickly.
After obtaining a candidate set, instead of scattering particles uniformly in the target scenario, we scatter particles in the positions of the candidate set, as the initialization stage.
The following modules work as in Section~\ref{sec:basiclocalizingmodule}.

\section{Experiment}
\label{sec:experiment}
We implement the prototype of Lightitude on Android OS, using unmodified Google Nexus 4 and Nexus 7 as the receiving devices.
We evaluate Lightitude in an office about $720 m^2$ with $39$ common fluorescent lamps, and a floor in the school library about $960 m^2$ with $100$ common fluorescent lamps, as shown in Fig~\ref{fig:8:c} and Fig~\ref{fig:8:d}.
Height of the lamps is $2.5$m and $2.7$m respectively in these two scenarios.
We obtain the floor plan and positions of the lamps in advance.

\begin{figure*}[t]
  \centering
  \subfloat[Office scenario]{
    \includegraphics[width=0.24\textwidth]{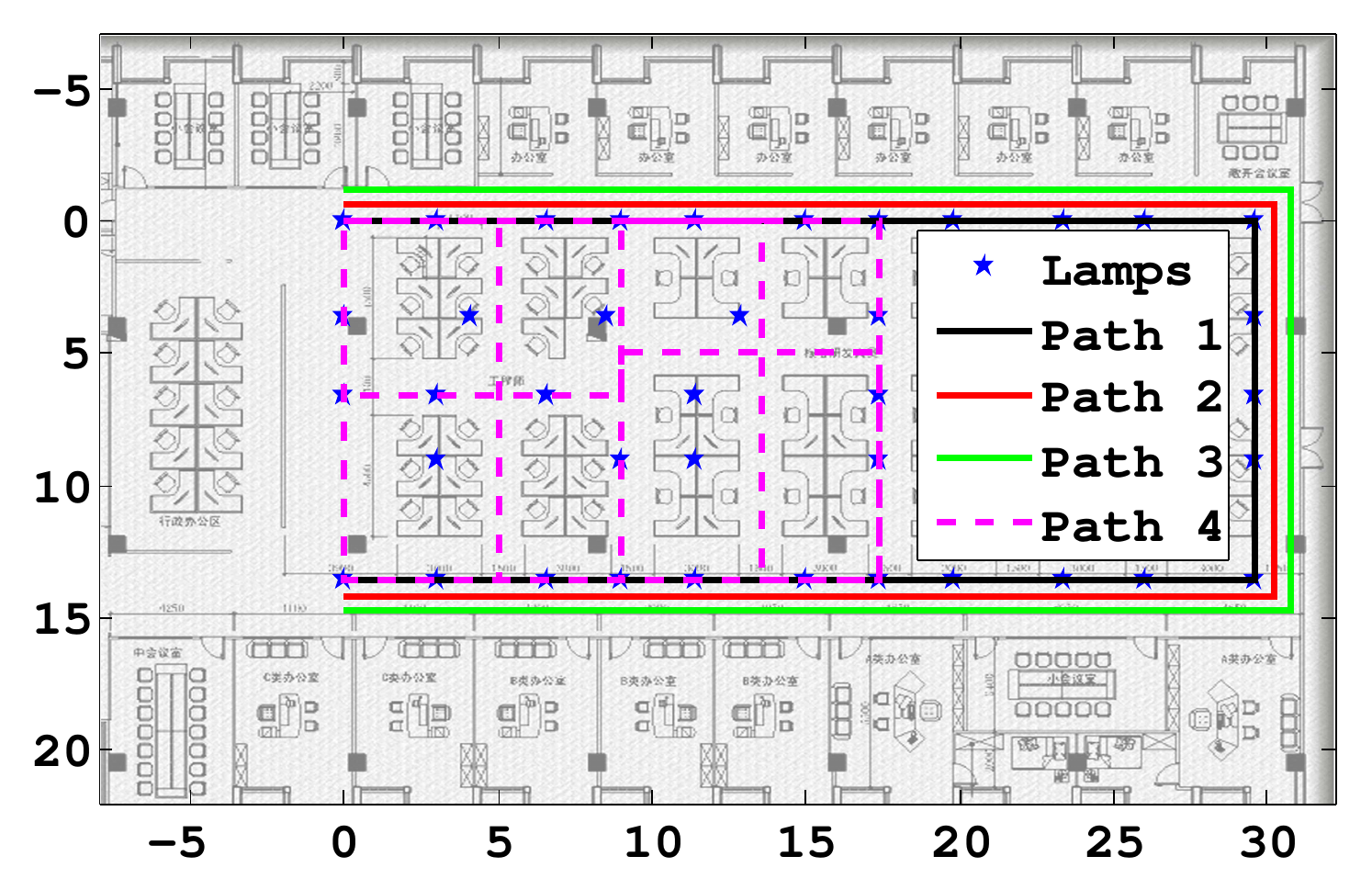}
    \label{fig:12:a}}
    \hfil
  \subfloat[Library scenario]{
    \includegraphics[width=0.24\textwidth]{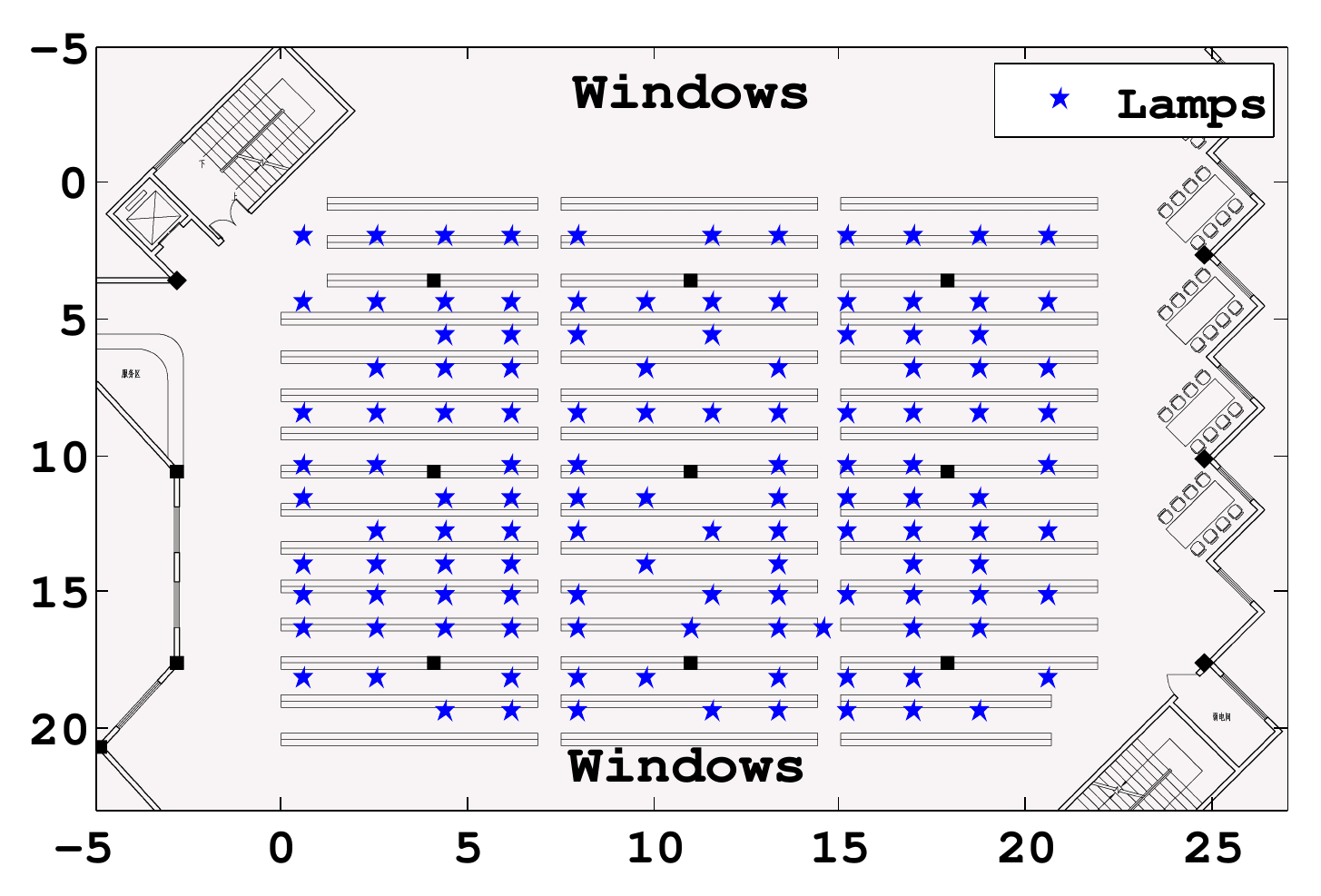}
    \label{fig:12:b}}
    \hfil
  \subfloat[Localizing benchmarks]{
    \includegraphics[width=0.2\textwidth]{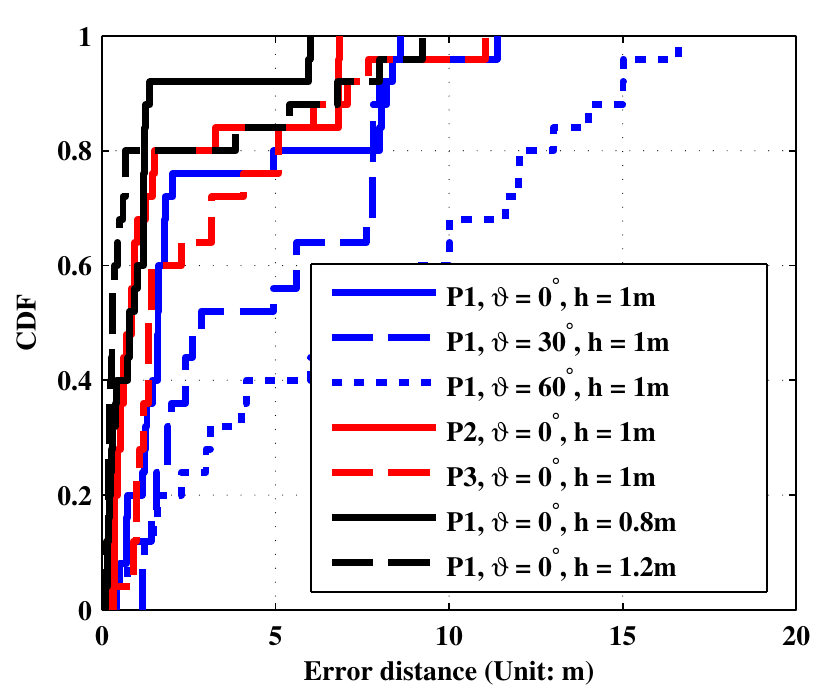}
    \label{fig:12:c}}
    \hfil
  \subfloat[Ideal RLS in target path]{
    \includegraphics[width=0.2\textwidth]{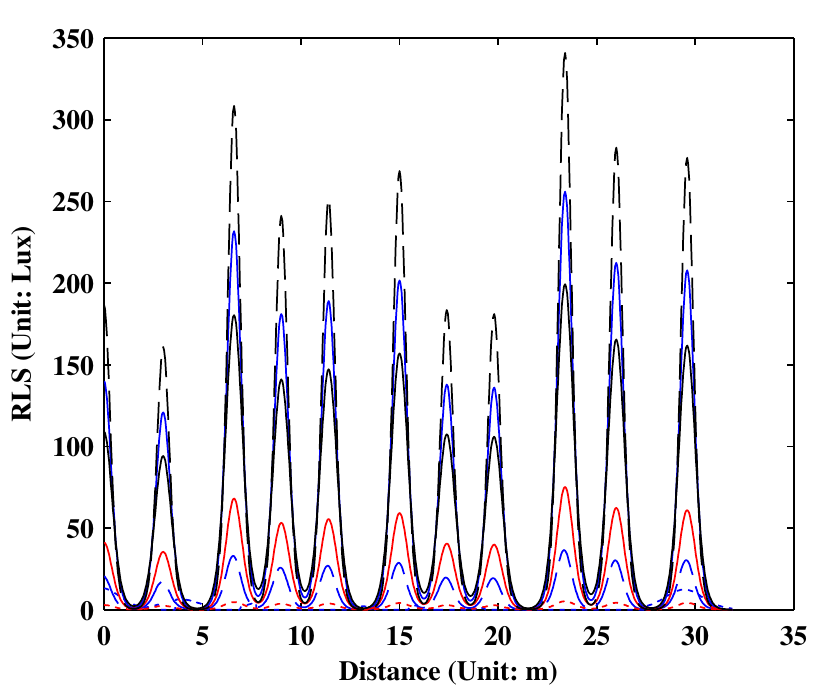}
     \label{fig:12:d}}
  \caption{(a): Experiments of \ref{exp:orientations} and \ref{exp:altitudes} are conducted in Path $1$. Experiment of \ref{exp:paths} is conducted in Path $1, 2, 3$. (b): Library scenario. Fig:c demonstrates the benchmarks of the localizing scheme in different conditions mentioned in~\ref{exp:localizingscheme}. (d) demonstrates the ideal RLS in different conditions mentioned in~\ref{exp:localizingscheme}. (c) and (d) share a same legend. In this legend, ``P1'' is short for Path 1, ``$1$m'' is short for height = $1$m.}
    \label{fig:12}
\end{figure*}

\subsection{Light Strength Model}
\label{exp:lightstrengthmodel}
We compare RLS collected by the device with RLS calculated by the light strength model to show the effectiveness of the model.
To obtain the calculation results of the light strength model from equation~\ref{eqn:multisources}, which further aggregate to form light strength distributions in Fig~\ref{fig:11:a} and Fig~\ref{fig:11:b}, we need to know the unique light strength parameter $L_i$ of each lamp.
For this purpose, volunteers bind the receiving device in the middle of a gyroscope, putting it at multiple locations and rotating it for a few seconds to collect RLS at arbitrary statuses of the device, as shown in Fig~\ref{fig:8:d}.
Integrating with the floor plan, we use equation~\ref{eqn:mse} to calculate the unique light strength parameter $L_i$ of each lamp. After that, we build the light strength model, and use it to calculate RLS for comparison with the RLS captured in the volunteer's site-survey.

Comparing with theoretical LCL and exponential rule, our light strength model has a minimum RMSE.
The error of RLS is limited within $10$ Lux in $50$ percentages and $50$ Lux in $95$ percentages, as shown in Fig~\ref{fig:12:c}.
Big deviation occurs mainly at $\theta$ ranging from $0^\circ$ to $30^\circ$, as shown in Fig~\ref{fig:12:d}.
It's because we use light incidence angle $\theta$ to cover changes in roll, pitch and yaw of the receiving device.
In reality, the position of the light sensor is not in the middle of the receiving device so rotation in pitch and yaw have a little different impact on RLS.
Moreover, from $\theta = 0^\circ$ to $\theta = 30^\circ$, RLS decays rapidly, and this rapid attenuation amplifies the difference.
We take this difference as the systematic error of the light strength model.

\subsection{Localizing Scheme: User influences RLS}
\label{exp:localizingscheme}
In this section, we validate the performance of the localizing scheme in different conditions of the user, in order to explore its applicability.
We divide the experiment into three parts, and each of them deals with an influencing factor of RLS: orientation of the receiving device (determines roll, pitch and yaw of the receiving device), user's route (determines receiving device's $2$D coordinate) and receiving device's height.

If without specific notification, in the following benchmarks, we set following parameters of Lightitude fixed: particle granularity $G$, number of stored particle $N$ at end of every iteration, search scope $S$ of the user's stride length and the smoothing factor $K$.
The particle scattering granularity $G$ is fixed at a particle per $0.1m^3$.
At the meantime, only $N = 100$ particles with the strongest weights are selected as the initial particles in each iteration, in order to reduce the overhead computation overhead of the localizing scheme.
We set the minimum stride length of user $r = 0.6$m and the maximum stride length of user $R = 1$m, so $S = R-r = 0.4m$ determines the width of the ring in Fig~\ref{fig:8:a}.
The smoothing factor $K$ is set fixed at $30$, which influences the convergence speed and positioning accuracy of the localizing scheme.
We'll discuss these influence factors in the next subsection.

\subsubsection{Device's Orientation}
\label{exp:orientations}
In different orientations of the device, multiple radiation / incidence angles exist according to different lamps.
But it's hard to control these angles explicitly in user's route, so we use the device's deviation angle from horizontal plane $\vartheta$ to represent the orientation of the device.
We define $\vartheta = 0^\circ$, if the facing orientation of the device is perpendicular to the horizontal plane.
In this experiment, we select $25$ positions in path $1$ as the starting points of the volunteer, as shown in Fig~\ref{fig:12:a}. These positions start from $(0, 0)$, and adjacent pairs of them is separated by $1$m.
Starting from these points, one volunteer walks toward the far end of the path.
The volunteer keeps the device at height $1$m roughly, with an average stride length $0.8$m and a constant walking speed.
We conduct experiments in $\vartheta = 0^\circ, 30^\circ, 60^\circ$ respectively, and the positioning accuracy is shown in Fig~\ref{fig:12:c}.
We notice that when the volunteer faces the device roughly up all the time ($\vartheta = 0^\circ$), the positioning error is limited within $1.60$m in $50$ percentages, and with a overall mean accuracy $2.95$m.
When the volunteer walks with $\vartheta$ around $30^\circ$, relatively big errors occur occasionally; when the volunteer walks with $\vartheta$ around $60^\circ$, the performance gets worse.

The crucial reason is that, with a big deviation angle of the device, RLS trend of the ground truth trace is weak, and has little difference compared with the RLS trend of other candidate traces.
Using the light strength model, we calculate the ideal light strength distribution in path $1$ with different $\vartheta = 0^\circ, 30^\circ, 60^\circ$ at height $1$m, as shown in Fig~\ref{fig:12:d}.
When $\vartheta = 0^\circ$, RLS trend of the ground truth trace is obviously distinguishable; when $\vartheta = 30^\circ$, the RLS trend is weak; when $\vartheta = 60^\circ$, the RLS trend is almost invisible.

\subsubsection{Different Paths}
\label{exp:paths}
Different routes of the user determine the relative distance from the device to the lamp set, which further influence collected RLS.
In this experiment, we select the same $25$ positions in path $1$, and their aligned positions in path $2$ and path $3$ as starting points, as shown in Fig~\ref{fig:12:a}.
Starting from these points, one volunteer walks toward the the far end of the paths.
The volunteer keeps $\vartheta = 0^\circ$ and height of device at $1$m roughly, with an average stride length $0.8$m and a constant walking speed.
As a result, the localizing module has a similar performance in these three paralleled paths, as shown in Fig~\ref{fig:12:c}.
However, there exist occasional errors, for the volunteer may start from points with inadequate lighting condition.
This generates a weak initial RLS that may mislead the initialization stage of the localizing module.
So the ground truth starting position (particle) may be eliminated at the very beginning.
This drawback is generated by a trade-off between computation overhead and positioning accuracy.
Because we cannot afford the computation overhead to process all particles in each iteration, only part of particles with strongest weights are chosen as starting particles in the next iteration.
A weak initial RLS together with the systematic error of the light strength model may eliminate the ground truth particle with a high probability.
To avoid this RLS ambiguity, the localizing scheme records only IMU information for back-propagation until it detects an obvious RLS.

\subsubsection{Device's altitudes}
\label{exp:altitudes}
In the user's route, one's movement gives rise to fluctuation of the device's altitudes, even the user tries to keep it stable.
Scattering particles to cover the full altitudes of the scenario is consumptive because users rarely put the device too high or too low.
To reduce the overall computation overhead, in Lightitude, we set a fixed altitudes range $0.8$m$\sim$$1.2$m for particle scattering, with step size $0.1$m.
This range shall be modified according to different smart devices. (e.g., if the user equips a smart-glass.)

In this experiment, we select the same $25$ positions in path $1$ as the starting points of the volunteer, as shown in Fig~\ref{fig:12:a}.
We conduct experiments in different heights $0.8$m, $1$m, $1.2$m respectively.
One volunteer starts from these points, and walks toward the far end of the path.
Volunteer keeps $\vartheta = 0^\circ$ roughly, with an average stride length $0.8$m and a constant walking speed.
The positioning accuracy is shown in Fig~\ref{fig:12:c}.
We notice that the performances in three different altitudes of the receiving device are similar, on account of the distinguishable RLS trend depicted in Fig~\ref{fig:12:d}.

\begin{figure*}[t]
  \centering
  \subfloat[Accuracy (granularity G)]{
    \includegraphics[width=0.22\textwidth]{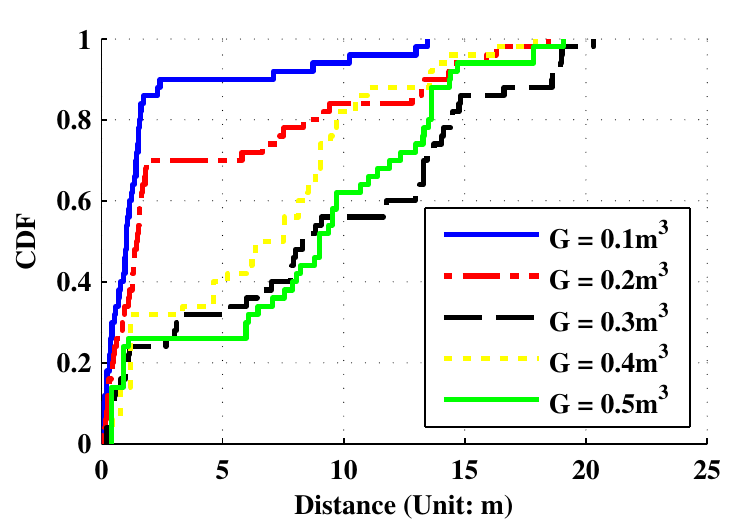}
    \label{fig:13:a}}
    \hfil
  \subfloat[Convergence speed (G)]{
    \includegraphics[width=0.22\textwidth]{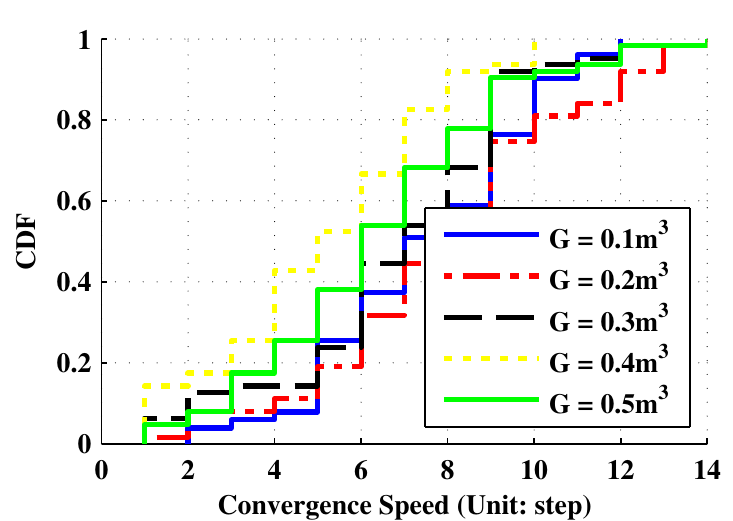}
    \label{fig:13:b}}
    \hfil
   \subfloat[Accuracy (particle number N)]{
    \includegraphics[width=0.22\textwidth]{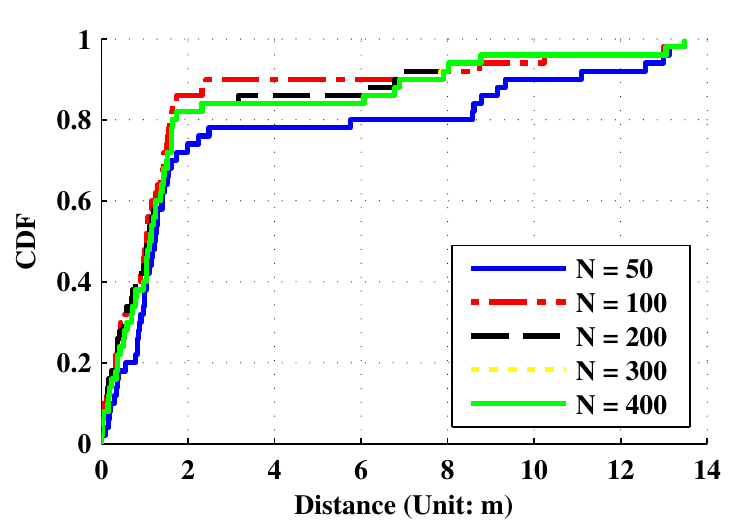}
    \label{fig:13:c}}
    \hfil
  \subfloat[Convergence speed (N)]{
    \includegraphics[width=0.22\textwidth]{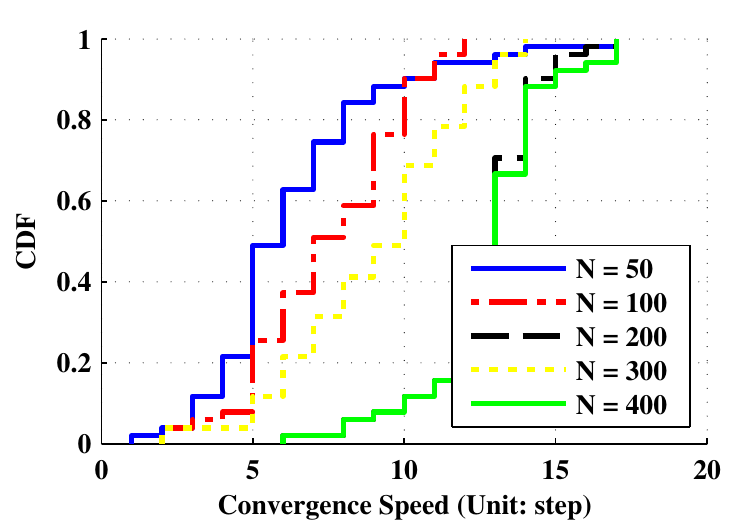}
    \label{fig:13:d}}
  \caption{Benchmarks on granularity of particles and particle number $N$.}
    \label{fig:13}
\end{figure*}

\subsection{Localizing Scheme: Parameters}
\label{exp:parameters}
In the above section, we validate the performance of the localizing scheme in different conditions of the user, under fixed parameters of the localizing scheme.
In this subsection, we'll discuss these parameters through comprehensive experiments.
We divide the experiment into four parts, and each of them deals with a parameter: particle granularity $G$, number of stored particle $N$ at end of every iteration, search scope $S$ of the user's stride length and the smoothing factor $K$.
If without specific notification, in the following benchmarks, we set the particle scattering granularity $G$ fixed at a particle per $0.1m^3$.
At the meantime, only $N = 100$ particles with the strongest weights are selected as the initial particles in each iteration, in order to reduce the overhead computation overhead of the localizing scheme.
We set the minimum stride length of user $r = 0.6$m and the maximum stride length of user $R = 1$m, so $S = R - r = 0.4m$ determines the width of the ring in Fig~\ref{fig:8:a}.
The smoothing factor $K$ is set fixed at $30$, which influence the convergence speed and positioning accuracy of the localizing scheme.

To make the benchmark result robust and scale, we evaluate Lightitude's performance in the library scenario, which is more severe than the office:
1) The lamps are denser so more lamps share similar light strength characteristics, which raises the mismatch possibility;
2) The gap between adjacent bookshelves is narrow (about $1$m), so accurately locating user between target bookshelves is a formidable task;
3) Long bookshelves generate shading effect, as shown in Fig~\ref{fig:6:d};
4) Large windows make interference of sunlight obvious.

\subsubsection{Granularity of particles $G$}
\label{exp:granularityofparticles}
A finer granularity of particles means when running Lightitude, more particles are scatter uniformly in the target scenario, meanwhile the granularity of ``children '' particle searching is fine-grained.
A finer granularity will make positioning result more accurate, but reduce the efficiency of the localizing module on the other hand.
In this experiment, we test the choice of $G$ on the performance of the localizing scheme.
We select an arbitrary $50$ positions in the library scenario as the starting points of the volunteer.
Starting from these points, one volunteer walks arbitrarily for $20$ steps to collect data, as the input of Lightitude.
The volunteer keeps the device in hand with a stable status, meanwhile with an average stride length $0.8$m and a constant walking speed.
We conduct experiments in granularity $G = 0.1, 0.2, 0.3, 0.4, 0.5m^3$ respectively, and the positioning accuracy is shown in Fig~\ref{fig:13:a}.
We notice that with the increment of $G$, the convergence speed becomes fast as shown in Fig~\ref{fig:13:b}, with the cost of loss in positioning accuracy.
The crucial reason behind this phenomenon is that, a coarse-grained $G$ makes particles sparse in the scenario.
As a result, particles with strong weight have a possibility not to be even initialized at the very beginning.

Facing this dilemma, we select $G = 0.1m^3$ to ensure robustness of Lightitude, which means Lightitude scatter particles in $0.1 \times 0.1 \times 0.1 m^3$.
In following overall experiments, we set $G = 0.1m^3$ as default.

\subsubsection{Particle number $N$}
\label{exp:particlenumber}
In the localizing scheme, we store a top $N$ particles with strongest weight as the candidate of the next iteration.
The reason why we not take all newly generated ones as candidates is giving the efficiency problem:
since storing more candidates makes the localizing scheme robust to occasional weight deviation of the ground truth particle, however, a big number of candidates cause a huge computation overhead.

In this experiment, we test the choice of $N$ on the performance of the localizing scheme.
We select the same $50$ positions in the library scenario as the starting points of the volunteer.
Starting from these points, one volunteer walks arbitrarily for $20$ steps to collect data, as the input of Lightitude.
The volunteer keeps the device in hand with a stable status, meanwhile with an average stride length $0.8$m and a constant walking speed.
We conduct experiments with particle number $N$ fixed at $50, 100, 200, 300, 400$ respectively, and the positioning accuracy is shown in Fig~\ref{fig:13:c}.
We notice that with the increment of $N$, the positioning accuracy is increasing.
But at the meantime, the convergence speed becomes slow, as shown in Fig~\ref{fig:13:d}.

In this paper, we select $N = 100$, using which Lightitude achieves satisfactory positioning result meanwhile has a relative fast converge speed.
In following overall experiments, we set $N = 100$ as default.

\begin{figure*}[t]
  \centering
  \subfloat[Accuracy (Search Range S)]{
    \includegraphics[width=0.22\textwidth]{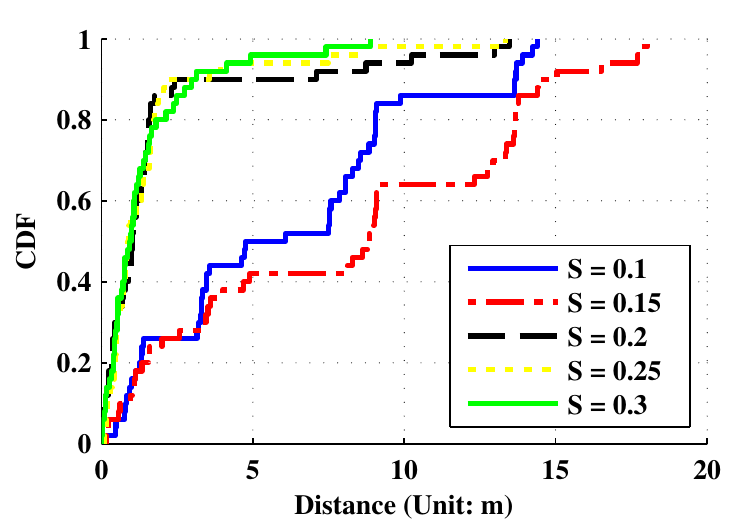}
    \label{fig:14:a}}
    \hfil
  \subfloat[Convergence speed (S)]{
    \includegraphics[width=0.22\textwidth]{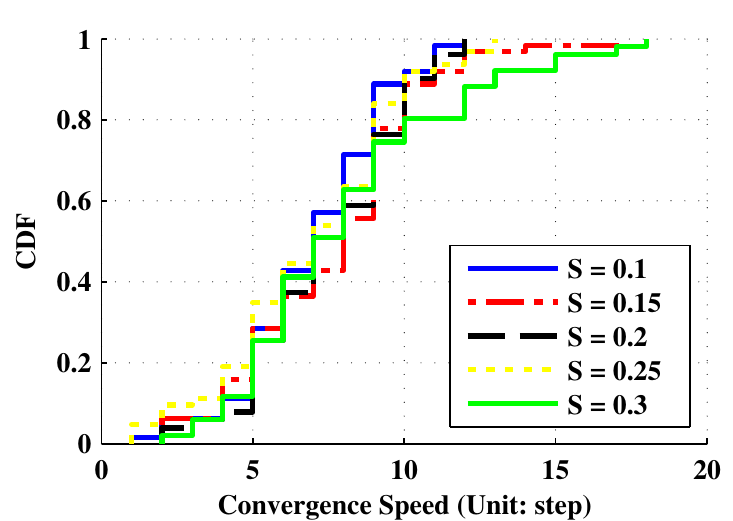}
    \label{fig:14:b}}
    \hfil
  \subfloat[Accuracy (Smoothing factor K)]{
    \includegraphics[width=0.22\textwidth]{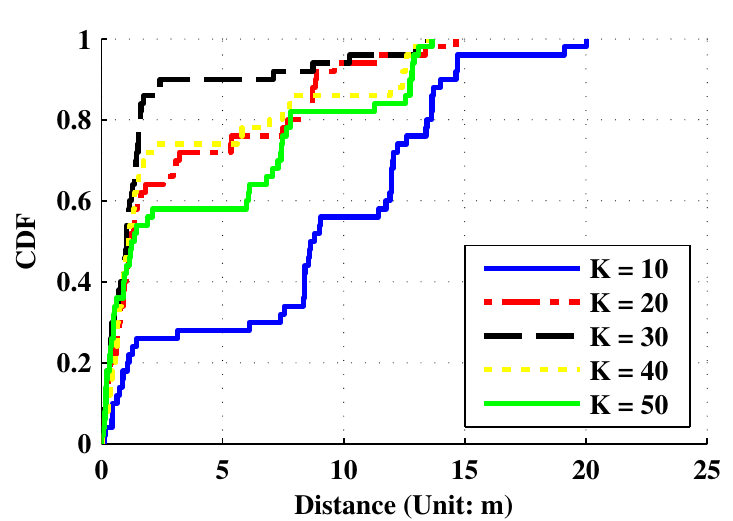}
    \label{fig:14:c}}
    \hfil
  \subfloat[Convergence speed (K)]{
    \includegraphics[width=0.22\textwidth]{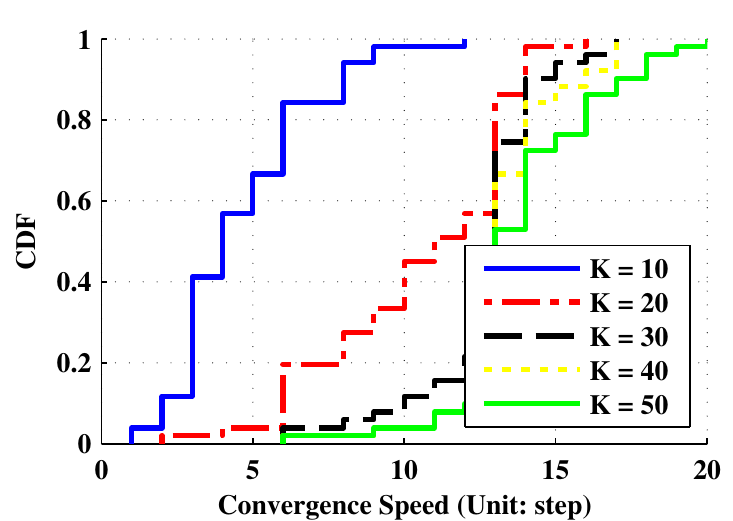}
    \label{fig:14:d}}
  \caption{Benchmarks on search scope of user's stride length and Smoothing Factor $K$.}
    \label{fig:14}
\end{figure*}

\subsubsection{Search scope of user's stride length $S$}
\label{exp:searchscope}
In Section~\ref{sec:localizingmodule}, we assume the user's minimum stride length is $r$ and maximum stride length is $R$, consequently a ring centering at the ``mother'' particles, ranging from $r$ to $R$ is the band for their ``children'' particles, as shown in Fig~\ref{fig:8:a}.
In this experiment, we test the choice of $S$ on the performance of the localizing scheme.
We test the choice of $S$ on the performance of the localizing scheme.
We select the same $50$ positions in the library scenario as the starting points of the volunteer.
Starting from these points, one volunteer walks arbitrarily for $20$ steps.
The volunteer keeps the device in hand with a stable status, meanwhile with a uniform walking speed.
Since the average stride length of a human being is about $30$ inches ($0.76m$)~\cite{meanpace}, we select $0.8$ as the center of the generated ring, and conduct experiments in radius of the ring $0.1 (r=0.7, R=0.9), 0.15 (r = 0.65, R = 0.95), 0.2 (r = 0.6, R = 1), 0.25 (r = 0.55, R = 1.05), 0.3 (r = 0.5, R = 1.1)$ respectively, and the positioning accuracy is shown in Fig~\ref{fig:14:a}.
We notice that a bigger $S$ makes positioning result more accurate, but the improvement is not significant.
The reason is that, human beings will rarely walk too fast or too slow upon using Lightitude, and their paces are around the average pace, which is the center of the search scope.
As a result, the ground truth particles are almost within $S$, even if $S$ is small.
At the meantime, a bigger $S$ reduces the convergence speed on the other hand.

In this paper, we select $S = 0.2$, using which Lightitude achieves satisfactory positioning result meanwhile has a relative fast converge speed.
In following overall experiments, we set $S = 0.2$ as default.

\subsubsection{Smoothing factor $K$}
\label{exp:smoothingfactor}
The smoothing factor $K$ in equation~\ref{eqn:particleweight} determines the weight of particles.
A bigger $K$ makes the localizing scheme robust to occasional deviation of the ground truth particle, but the localizing scheme will take more time to reach the convergence condition, since a bigger $K$ prevent the particles from obtaining big weights quickly, thus reduce the convergence speed of the localizing scheme.

In this experiment, we test the choice of $K$ on the performance of the localizing scheme.
We select the same $50$ positions in the library scenario as the starting points of the volunteer, as shown in Fig~\ref{fig:12:b}.
Starting from these points, one volunteer walks arbitrarily for $20$ steps to collect data, as the input of Lightitude.
The volunteer keeps the device in hand with a stable status, meanwhile with an average stride length $0.8$m and a constant walking speed.
We conduct experiments in smoothing factor $K = 10, 20, 30, 40, 50$ respectively, and the positioning accuracy is shown in Fig~\ref{fig:14:c}.
We notice that, in these smoothing factors, $K = 30$ provide the robust positioning result, meanwhile the convergence speed is relatively fast.
As a consequence, In following overall experiments, we set $K = 30$ as default.

\subsection{Putting It All Together}
\label{exp:puttingitalltogether}
Taking all influence factors together, we evaluate Lightitude's performance in different scenarios.

\subsubsection{Pure Light}
\label{exp:purelight}
\begin{figure}[b]
  \centering
  \subfloat[Accuracies (Length)]{
    \includegraphics[width=0.23\textwidth]{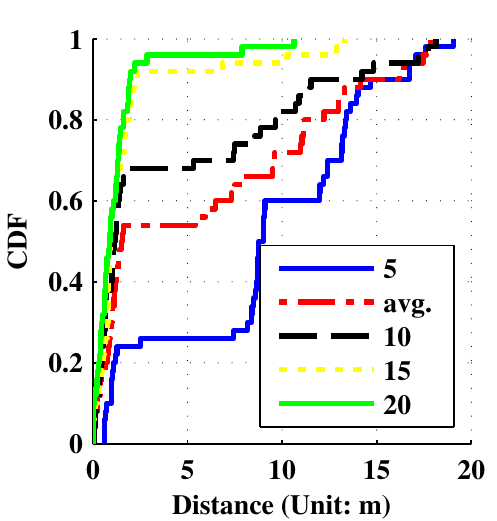}
    \label{fig:16:a}}
    \hfil
  \subfloat[Step error (Length)]{
    \includegraphics[width=0.23\textwidth]{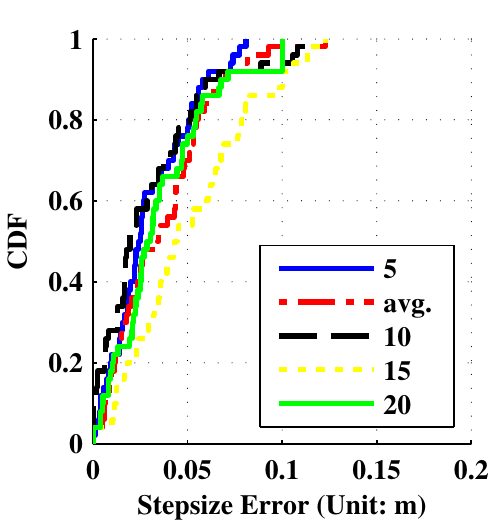}
    \label{fig:16:b}}
  \caption{Positioning Accuracy in different lengths of traces.}
    \label{fig:16}
\end{figure}

\begin{figure*}[t]
  \centering
  \subfloat[Positioning accuracies]{
    \includegraphics[width=0.23\textwidth]{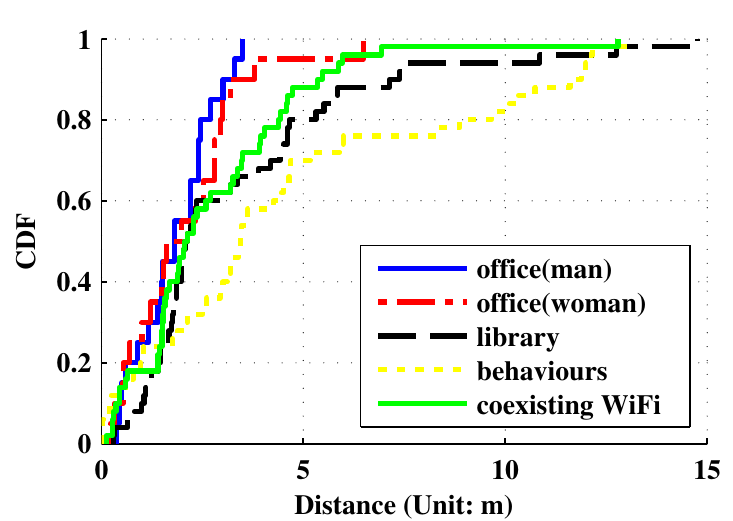}
    \label{fig:15:a}}
    \hfil
  \subfloat[Convergence speeds]{
    \includegraphics[width=0.23\textwidth]{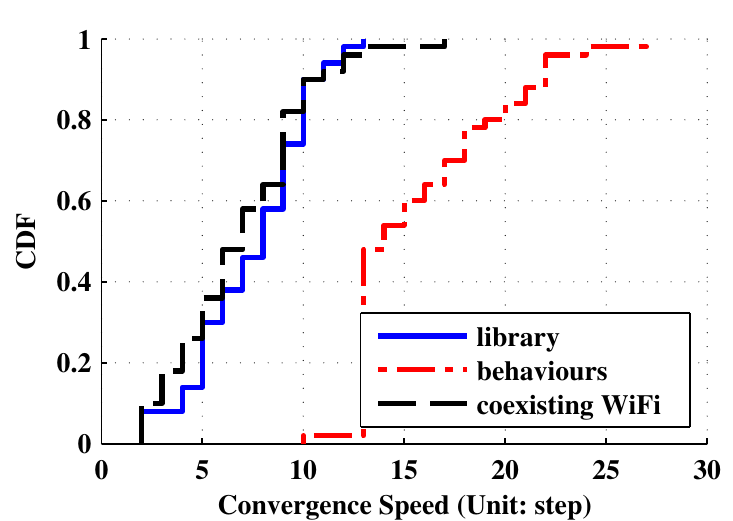}
    \label{fig:15:b}}
    \hfil
  \subfloat[Stepsize errors]{
    \includegraphics[width=0.23\textwidth]{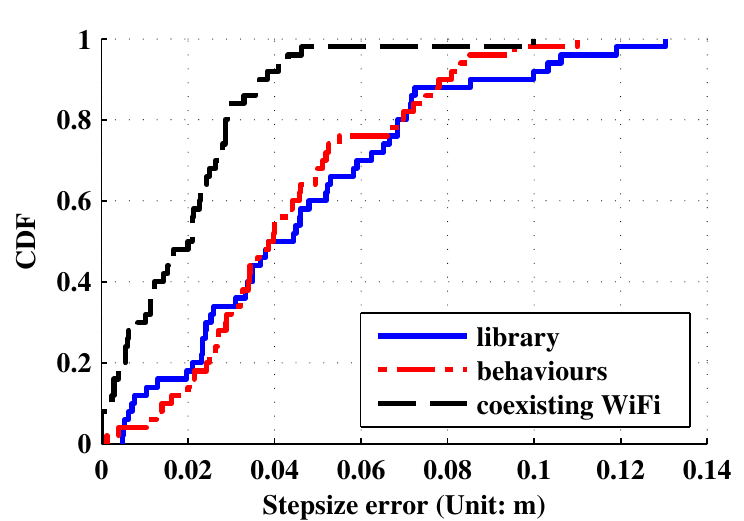}
    \label{fig:15:c}}
    \hfil
  \subfloat[Robustness of Lightitude]{
    \includegraphics[width=0.23\textwidth]{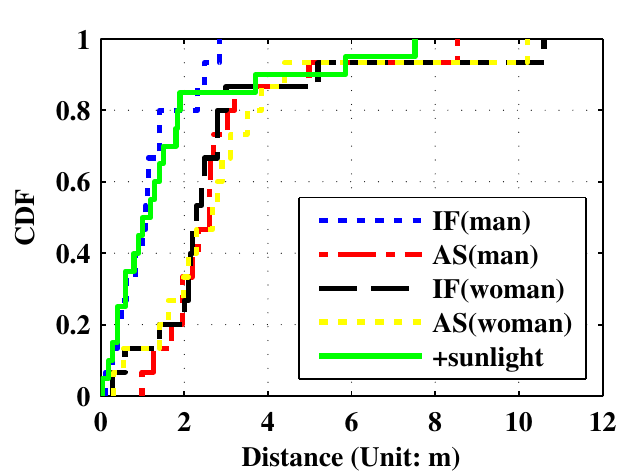}
    \label{fig:15:d}}
  \caption{Positioning Accuracy of pure light, with user's unpredictable behaviours and coexisting WiFi.
  IF is short for ``In front of the user's body'', meanwhile AS is short for ``At the side of the user's body''. ``$+$sunlight'' means volunteers start from positions which are influenced by sunlight.}
    \label{fig:15}
\end{figure*}

We first validate the performance of pure Lightitude, and then design schemes to ensure its robustness and speed it convergence speed.
In the office scenario, two volunteers, one man with average stride length $0.8$m and one woman with average stride length $0.5$m are recruited to validate Lightitude's performance with volunteer diversity.
We randomly select $20$ positions in path $1, 2, 3, 4$ in Fig~\ref{fig:12:a} as the starting points of the volunteers.
The results are shown in Fig~\ref{fig:15:a}.
Lightitude achieves mean accuracy $1.93$m in the office scenario.

In the library scenario, we recruit the same two volunteers to validate Lightitude's performance.
We randomly select $50$ positions in the scenario as the starting points of the volunteers.
The results are shown in Fig~\ref{fig:15:a}.
Lightitude achieves mean accuracy $3.42$m, slightly worse than that in the office because occasional mismatch happens.
The user need to walk across for average $7.42$ steps for positioning, which means the user need to walk across only on average $3$ lamps.

\subsubsection{Localizing continuously}
\label{exp:localizingcontinuously}
One thing needed to be noted is that, comparing with WiFi-based positioning schemes like RADAR~\cite{bahl2000radar}, whose CDF curve is quite smooth, CDF of Lightitude is zigzag, especially in the tail.
The reason is that, candidate results of WiFi-based schemes are limited in restrict area, so big errors rarely happens: at least the result is not far from the ground truth one.
\textit{On the contrary, Lightitude is caught in a win or go-home dilemma: if Lightitude successfully locate user, the precision is quite satisfactory; or big error happens, since there exist no locality in the candidate set.}
Basing on this observation, a nature question arise: if the user walks longer, will Lightitude performs better?

We evaluate the performance of Lightitude by using traces with different lengths provided by the user.
We randomly select $50$ positions in the scenario as the starting points of the volunteers.
Starting from these points, the user walks traces with different lengths: $5$ steps, $8$ steps (average convergence speed), $10$ steps, $15$ steps, $20$ steps.
The results are shown in Fig~\ref{fig:16:a} and Fig~\ref{fig:16:b}.
Under different lengths, the positioning accuracy is $8.88$m, $5.75$m, $4.20$m, $1.79$m, $1.26$m respectively.
We notice that, when the user walks a longer distance, the positioning accuracy is increasing.
The reason is that, in big scenarios with dense light sources, RLS trend in several steps is not distinguishable enough to uniquely locate user.
Since the user doesn't know when Lightitude converges by leveraging the convergence conditions, it is suggested to walk longer and using Lightitude simultaneously to achieve a higher positioning accuracy.

\subsubsection{Preventing module}
\label{exp:preventingmodule}
The normal localization module has a risk to be misled by user's unpredictable activities.
Facing these dilemmas, we design a prevention module for Lightitude, in order to increase the robustness of Lightitude even facing with pre-mentioned unpredictable behaviours.
we recruit the same two volunteers to validate Lightitude's performance.
We randomly select $50$ positions in the scenario as the starting points of the volunteers.
Upon user's walking, one puts the device in the pocket, and the put it out again after walking for $4$ steps.
The positioning accuracy is shown in Fig~\ref{fig:15:a}.
With the unpredictable behaviours of the user, Lightitude still can precisely locate user in target shelves in $30$ out of $50$ times.

\subsubsection{Coexisting with WiFi}
\label{exp:coexistingwifi}
In this experiment, we first use a WiFi fingerprint-based module once to provide a candidate position set, then exploit the light-based positioning module to prune this set.
We deploy a total $4$ APs, each in a corner of the scenario, and we sample a fingerprint near each lamp.
As a result, the fingerprinting granularity is about one fingerprint per $2.4 \times 2.4m^2$.
we recruit the same two volunteers to validate Lightitude's performance.
We randomly select $50$ positions in the scenario as the starting points of the volunteers.
The positioning accuracy is shown in Fig~\ref{fig:15:a}.
We notice that with the help of WiFi module, the positioning accuracy increases and yields mean accuracy $1.98$m, with the $74$-percentile results limited in $2$m.
The convergence speed is increasing as shown in~\ref{fig:15:b}.
Basing on this observation, Lightitude can be applied as an auxiliary subsystem coexisting with WiFi-based schemes, in order to provide a finer grained positioning service to the user.

\subsubsection{Sunlight/Shading by human's body}
\label{exp:sunlight}
In this experiment, we evaluate the performance of Lightitude under the interference of sunlight and the shielding effect of human body.
We recruit the same two volunteers to validate Lightitude's performance.
$20$ positions near the windows (at noon to get the interference of sunlight) are selected as the start points of the user.
We notice that when the user starts from positions with sunlight interference, Lightitude performs even better, and achieves mean accuracy $1.67$m.
The reason is that, sunlight interference provides Lightitude hints in particle scattering (near the windows), which decreases the possibility of mismatch.
However, together with the prevention module, when the user walks across the sunlight-interference area, the step number needed for convergence is increasing, which is determined by the size of the sunlight interference area.

We also evaluate Lightitude's performance under different human body's shading effects.
The same two volunteers conduct $20$ tests in two device-holding gestures respectively to get human-body's shading effect:
1) put the phone in front of the body;
2) put the phone at the side of the body.
As shown in Fig~\ref{fig:9:b}, the performance is similar in different gestures of the user, so the shading effect does not have a strong impact.
To conclude, Lightitude is robust even facing with the sunlight interference and shading effect of human body.

In total, by exploiting merely ubiquitous visible lights and COTS device, Lightitude achieves mean accuracy $1.93$m in the office scenario, and achieves mean accuracy $3.42$m in the library under different conditions.
The errors are due to the path similarity in a big scenario, and can be solved by coexisting with schemes basing on other mediums (e.g., WiFi, magnetic field).
Besides, Lightitude exploits sunlight interference which was supposed to be a obstacle, and achieves mean accuracy $1.67$m in this scenario.
Lightitude's robustness ensures that it performs satisfactorily even under shading of human-body and unpredictable behaviours of users.

Lightitude faces the same limitation with state-of-the-art visible light based positioning approaches~\cite{liepsilon,luxapose2014}, where users need to expose their receiving devices with occasional ideal status to ensure better positioning result.
However, there is not such a problem for smart-glass users and smart-watch users.
For smart-phone or smart-watch, it is essential for the users to take out these devices and operate for positioning, which exposes the devices to the lights and trigger Lightitude.
We validate this by using smart watch Moto 360 as an receiving device of Lightitude.

\section{Conclusion}
\label{sec:conclusion}
In this paper, we propose Lightitude, an indoor localization scheme exploiting only ubiquitous visible lights.
We have identified and overcome two technical challenges. First, we propose and validate a light strength model to avoid frequent site-survey and database maintenance.
Second, we harness user's mobility to generate spatial-related RLS to tackle the position ambiguity problem of a single RLS. Our evaluation in typical office and library environments confirms the effectiveness and the robustness of Lightitude.
Due to the ubiquity and zero-cost of Lightitude, it can be directly applied as an auxiliary subsystem coexisting with WiFi-based schemes or even an independent positioning system.
In our future work, we will combine Lightitude with schemes using other existing infrastructures, like magnetic field~\cite{zheng2014travi}, to improve the accuracy and robustness.

\bibliographystyle{plain}
\bibliography{lightitude-jversion}

\begin{thebibliography}{10}

\bibitem{lcl}
http://en.wikipedia.org/wiki/Lambert's\_cosine\_law.

\bibitem{connolly2013}
P. connolly and d. boone. indoor location in retail: Where is the money? abi
  research report, 2013.

\bibitem{meanpace}
Pace (unit).
\newblock http://en.wikipedia.org/wiki/Pace\_(unit).

\bibitem{adib20133d}
Fadel Adib, Zach Kabelac, Dina Katabi, and Robert~C Miller.
\newblock 3d tracking via body radio reflections.
\newblock In {\em Usenix NSDI}, volume~14, 2013.

\bibitem{adib2013see}
Fadel Adib and Dina Katabi.
\newblock See through walls with wifi!
\newblock In {\em SIGCOMM}, pages 75--86. ACM, 2013.

\bibitem{bahl2000radar}
Paramvir Bahl and Venkata~N Padmanabhan.
\newblock Radar: An in-building rf-based user location and tracking system.
\newblock In {\em INFOCOM}, volume~2, pages 775--784. IEEE, 2000.

\bibitem{chen2012fm}
Yin Chen, Dimitrios Lymberopoulos, Jie Liu, and Bodhi Priyantha.
\newblock Fm-based indoor localization.
\newblock In {\em MobiSys}, pages 169--182. ACM, 2012.

\bibitem{DBLP:conf/infocom/HuangXLLMYL14}
Wenchao Huang, Yan Xiong, Xiang-Yang Li, Hao Lin, XuFei Mao, Panlong Yang, and
  Yunhao Liu.
\newblock Shake and walk: Acoustic direction finding and fine-grained indoor
  localization using smartphones.
\newblock In {\em INFOCOM, 2014 Proceedings IEEE}, pages 370--378. IEEE, 2014.

\bibitem{jimenez2009comparison}
AR~Jimenez, F~Seco, C~Prieto, and J~Guevara.
\newblock A comparison of pedestrian dead-reckoning algorithms using a low-cost
  mems imu.
\newblock In {\em Intelligent Signal Processing, 2009. WISP 2009. IEEE
  International Symposium on}, pages 37--42. IEEE, 2009.

\bibitem{joshi2013pinpoint}
Kiran Joshi, Steven Hong, and Sachin Katti.
\newblock Pinpoint: Localizing interfering radios.
\newblock In {\em USENIX NSDI}, 2013.

\bibitem{keogh_exact_2005}
Eamonn Keogh and Chotirat~Ann Ratanamahatana.
\newblock Exact indexing of dynamic time warping.
\newblock {\em Knowledge and Information Systems}, 7(3):358\textendash{}386,
  March 2005.

\bibitem{liepsilon}
Liqun Li, Pan Hu, Chunyi Peng, Guobin Shen, and Feng Zhao.
\newblock Epsilon: a visible light based positioning system.
\newblock In {\em Proceedings of the 11th USENIX Conference on NSDI}, pages
  331--343, 2014.

\bibitem{liu2012push}
Hongbo Liu, Yu~Gan, Jie Yang, Simon Sidhom, Yan Wang, Yingying Chen, and Fan
  Ye.
\newblock Push the limit of wifi based localization for smartphones.
\newblock In {\em MOBICOM}, pages 305--316. ACM, 2012.

\bibitem{liu2008basic}
Xiaohan Liu, Hideo Makino, and Yoshinobu Maeda.
\newblock Basic study on indoor location estimation using visible light
  communication platform.
\newblock In {\em Engineering in Medicine and Biology Society, 2008. EMBS 2008.
  30th Annual International Conference}, pages 2377--2380. IEEE, 2008.

\bibitem{peng2007beepbeep}
Chunyi Peng, Guobin Shen, Yongguang Zhang, Yanlin Li, and Kun Tan.
\newblock Beepbeep: a high accuracy acoustic ranging system using cots mobile
  devices.
\newblock In {\em Proceedings of the 5th international conference on Embedded
  networked sensor systems}, pages 1--14. ACM, 2007.

\bibitem{rai2012zee}
Anshul Rai, Krishna~Kant Chintalapudi, Venkata~N Padmanabhan, and Rijurekha
  Sen.
\newblock Zee: Zero-effort crowdsourcing for indoor localization.
\newblock In {\em MOBICOM}, pages 293--304. ACM, 2012.

\bibitem{rajagopal2014visual}
Niranjini Rajagopal, Patrick Lazik, and Anthony Rowe.
\newblock Visual light landmarks for mobile devices.
\newblock In {\em IPSN}, pages 249--260, 2014.

\bibitem{randall2007luxtrace}
Julian Randall, Oliver Amft, J{\"u}rgen Bohn, and Martin Burri.
\newblock Luxtrace: indoor positioning using building illumination.
\newblock {\em Personal and ubiquitous computing}, 11(6):417--428, 2007.

\bibitem{5433479}
Nishkam Ravi and Liviu Iftode.
\newblock {FiatLux: Fingerprinting Rooms Using Light Intensity}.

\bibitem{shen2013walkie}
Guobin Shen, Zhuo Chen, Peichao Zhang, Thomas Moscibroda, and Yongguang Zhang.
\newblock Walkie-markie: indoor pathway mapping made easy.
\newblock In {\em Proceedings of the 10th USENIX conference on Networked
  Systems Design and Implementation}, pages 85--98. USENIX Association, 2013.

\bibitem{wang2012no}
He~Wang, Souvik Sen, Ahmed Elgohary, Moustafa Farid, Moustafa Youssef, and
  Romit~Roy Choudhury.
\newblock No need to war-drive: Unsupervised indoor localization.
\newblock In {\em MobiSys}, pages 197--210. ACM, 2012.

\bibitem{wang2013dude}
Jue Wang and Dina Katabi.
\newblock Dude, where's my card?: Rfid positioning that works with multipath
  and non-line of sight.
\newblock In {\em SIGCOMM}, 2013.

\bibitem{yang2012locating}
Zheng Yang, Chenshu Wu, and Yunhao Liu.
\newblock Locating in fingerprint space: wireless indoor localization with
  little human intervention.
\newblock In {\em MOBICOM}, pages 269--280. ACM, 2012.

\bibitem{luxapose2014}
Ko-Jen~Hsiao Ye-Sheng~Kuo, Pat~Pannuto and Prabal Dutta.
\newblock Luxapose: Indoor positioning with mobile phones and visible light.
\newblock In {\em MOBICOM}. ACM, 2014.

\bibitem{yoshino2008high}
Masaki Yoshino, Shinichiro Haruyama, and Masao Nakagawa.
\newblock High-accuracy positioning system using visible led lights and image
  sensor.
\newblock In {\em Radio and Wireless Symposium}, pages 439--442. IEEE, 2008.

\bibitem{youssef2005horus}
Moustafa Youssef and Ashok Agrawala.
\newblock The horus wlan location determination system.
\newblock In {\em MobiSys}, pages 205--218. ACM, 2005.

\bibitem{zheng2014travi}
Yuanqing Zheng, Guobin Shen, Liqun Li, Chunshui Zhao, Mo~Li, and Feng Zhao.
\newblock Travi-navi: Self-deployable indoor navigation system.
\newblock In {\em MOBICOM}, pages 471--482. ACM, 2014.

\end{thebibliography}


\end{document}